\documentclass[reprint,aps,floatfix,,prb, superscriptaddress,10pt]{revtex4-2}

\usepackage{setspace}

\usepackage{mlmodern}

\usepackage{calc}
\usepackage{amsmath,bm}
\usepackage{amssymb}
\usepackage{amsfonts}
\usepackage{mathrsfs}
\usepackage{mathtools}
\usepackage{csquotes}

\allowdisplaybreaks
\makeatletter
  \def\my@tag@font{\normalsize}
  \def\maketag@@@#1{\hbox{\m@th\normalfont\my@tag@font#1}}
  \let\amsmath@eqref\eqref
  \renewcommand\eqref[1]{{\let\my@tag@font\relax\amsmath@eqref{#1}}}
\makeatother

\usepackage{psfrag}
\usepackage{graphicx}
\usepackage[caption=false]{subfig}
\graphicspath{{Figures/}}
\usepackage{color,soul}
\usepackage[svgnames,x11names]{xcolor}
\usepackage{scalerel} 

\definecolor{Custom}{rgb}{0.,0.,0.8}
\usepackage{tabularx}

\usepackage[unicode,linktocpage=true]{hyperref}
\hypersetup{colorlinks=true,linkcolor=Custom,urlcolor =Custom,citecolor=Custom,anchorcolor=Custom} 

\usepackage[center]{titlesec} 

\titlespacing{\section}{10pt}{10pt}{10pt}
\titlespacing{\subsection}{10pt}{10pt}{10pt}
\titlespacing{\subsubsection}{10pt}{10pt}{10pt}

\titleformat{\section}[block]{\centering\bfseries}{\thesection.}{0.7em}{\MakeUppercase}
\titleformat{\subsection}[block]{\centering\bfseries}{\thesubsection.}{0.6em}{}
\titleformat{\subsubsection}[block]{\centering\itshape\bfseries\small}{\thesubsubsection.}{0.5em}{\small}

\def\nn{\nonumber}
\def\bs{\boldsymbol}
\newcommand{\mycomment}[1]{}

\begin{document}

\title{Topological pairing of composite fermions via criticality}

\author{N. Ne\v{s}kovi\'c}
\author{I. Vasi\'c}
\author{M.V. Milovanovi\'c}
\affiliation{Scientific Computing Laboratory, Center for the Study of Complex Systems,Institute of Physics Belgrade, University of Belgrade, Pregrevica 118, 11080 Belgrade, Serbia}


\begin{abstract}
 {\begin{spacing}{1.125} The fractional quantum Hall effect (FQHE) at the filling factor with an even denominator, $\frac{5}{2}$, occurs despite the expectation, due to the electron statistics, that the denominator must be an odd number. It is believed that the Cooper pairing of underlying quasiparticles, composite fermions (CFs), leads to the explanation of this effect. Such a state should have a Pfaffian form of the BCS wave function (due to the fully-polarized spin) and non-Abelian statistics of possible vortex-like excitations (due to the $p$-wave nature of the pairing). Here we expose the origin of pairing by using the effective dipole representation of the problem and show that pairing  is encoded in a Hamiltonian that describes the interaction of the charge density with dipoles i.e. the current of CFs. The necessary condition for the paired state to exist is the effective dipole physics at the Fermi level as a consequence of the non-trivial topology of the ideal band in which electrons live - a Landau level (LL); the paired state is a resolution of the unstable, critical behavior characterized by the distancing of correlation hole with respect to electron (and thus dipole) at the Fermi level due to the topology. We describe analytically this deconfined critical point, at which deconfinement of Majorana neutral fermions takes place. In the presence of large, short-range repulsive interaction inside a LL, the critical behavior may be stabilized  into a regularized Fermi-liquid-like (FLL) state, like the one that characterizes the physics in the lowest LL (LLL), but in general, for an interaction with slowly decaying pseudopotentials, the system is prone to pairing.\end{spacing}}

\end{abstract}

\maketitle

\pagestyle{plain}


\section{Introduction}

The fractional quantum Hall effect (FQHE) \cite{PhysRevLett.48.1559}  can be described as the existence, in the experimental data, of the plateaus of  the fractionally quantized Hall conductance with simultaneous minima in the longitudinal conductance of effectively  two-dimensional systems of electrons. This phenomenon occurs around special commensuration points. At the center of a plateau, the system is at a special commensuration point, for which the ratio of a specific volume for the orbital motion of an electron in the magnetic field, perpendicular to the system, with respect to the volume per particle in the system, is a simple fraction. The interaction effects are the most pronounced at these special points and lead to gapped, very stable liquid states of the system.

The FQHE at filling factor 5/2 \cite{PhysRevLett.59.1776} came  as a surprise because of the expectation, due to the electron fermionic statistics, that the denominator of the fraction should be an odd number. For the explanation of this effect, a proposal \cite{MOORE1991362}  was made for the state of spinless (spin-frozen) electrons, in a half-filled second Landau level (LL), that represents a pairing of underlying quasiparticles, composites of electrons and their correlation holes, known as composite fermions. The pairing in this well-known Moore-Read or Pfaffian state is a $p$-wave topological pairing of spinless fermions, and is a source of an unusual neutral fermion physics: Majorana physics in vortices and at the edges of the system. This leads to the non-Abelian statistics of the vortex, i.e., quasiparticle excitations, and potential use for the (topological) quantum computation \cite{RevModPhys.80.1083}.

A reason for the pairing of composite electrons (fermions) certainly lies in the constrained dynamics:  quantized  orbital motion of electrons in the (second) LL. The application of an effective projection of the physics into a single LL, is a way that most of the FQHE states are understood, i.e., by solving a purely interacting problem inside a portion (a single LL) of the complete Hilbert space. Other portions, LLs, are excluded due to the usually high cyclotron energy necessary for the inclusion of other LLs. Nevertheless, a complete understanding of the mechanism of what we may call a topological pairing (inside a LL, i.e., an ideal Chern band) is missing.

In this work we will discuss the mechanism for the topological pairing - the pairing inside the system with repulsive interactions, that is defined in an isolated (ideal) Chern band, i.e., a band with an exceptional connectedness.  The connectedness is exemplified in the noncommuting algebra (Girvin-MacDonald-Platzman (GMP) algebra) of density operators defined (projected into) the band.  The non-trivial topology of the band will force any {\em fermionic} quasiparticle composites [like composite
fermions (CFs)], effectively an overall neutral objects with two opposite charges,  i.e., dipoles of elementary particle and correlation hole, to have momentum proportional to its dipole moment (in the magnetic field as the cause of the topology) \cite{Read_1994,Read_1996}. Thus, at the Fermi momentum, the correlation hole is necessarily distanced from the elementary particle (electron), they avoid each other (which is unfavorable for the minimization of the interaction energy), and a Fermi-liquid (FL) state is unstable. Thus, instead of a FL state, we have a critical state, more precisely, a high energy, enriched with a discrete symmetry state, which decays, by spontaneously choosing one of  two possible symmetry-broken states that are paired states.

We will illustrate this mechanism on two well-known systems: (a) a system of bosons at filling factor one, in the lowest LL (LLL), and (b) a system of electrons that fill half of the available states in a LL (i.e., a system at filling factor 1/2) that, for certain interactions, will have the Pfaffian state as a ground state. Pfaffian physics is demonstrated usually on a sphere for a special shift, which acts as a bias for the Pfaffian state and, in our interpretation, as a small perturbing parameter which will break the high-energy discrete symmetry of the system and stabilize the system into one of possible two paired states of composite quasiparticles.

Common to both systems is the existence, or possible interpretation in terms, of (overall neutral) composite fermionic quasiparticles. Correlation holes in these systems are well defined, due to a sufficient incompressibility (hard to compress feature) of the systems, that enable correlation holes to become excitations with well-defined charge and statistics (at least) in the long-wavelength limit. Therefore, the composite : electron (boson) + correlation hole is well defined. The distancing of the correlation hole with respect to elementary electron (boson),  (so that the density of correlation holes becomes the density of real holes in the electron system, and thus energetically unfavorable for the usual repulsive  interactions),  necessarily accompanies the emergence of a mass and what we may call,  a critical Fermi-liquid-like (FLL) state as a prerequisite for the pairing.

The critical state may also decay into a regularized, stable state that represents a FLL state as in the case of electrons in the LLL. This happens because, for certain large short-range  repulsive interactions (for which dominant is the first Haldane pseudopotential in the case of half-filled LL of electrons), the distancing  of correlation hole with respect to electron is energetically favorable; real holes surround electron due to the interaction and, at the special half-filling condition, due to the particle-hole (PH) symmetry, correlation holes can be identified with real holes. In general, for  a gradually decaying pseudopotentials, in the case of electrons, and, without this  specific restriction, in the case of bosons, we  expect pairing (if neutral quasiparticles such as CFs can be identified in the system, which, we know, in the case of half-filled isolated LLs of electrons, or completely filled LLs of bosons). Thus, (liquid) phases with pairing should prevail in these systems ( if we exclude the possibility of broken translation symmetry states).

An extensive list of references on the physics of electrons at filling factor 5/2 can be found in Ref. \cite{MA2024324}. In Ref. \cite{scarola2000} a Cooper problem for composite fermions was considered.
Recently, in Refs. \cite{Sharma_2021,Sharma_2023,Sharma_2024}, an approach for half-filled LL electron systems, based on variational BCS FQHE wave function, was elaborated.

Section II is devoted to the system of bosons. In this section, the necessary formalism is introduced, and the mechanism for pairing with the critical behavior is explained. The Section III describes the physics of the electron system(s) and their pairing instabilities. The last section, Section IV, is devoted to conclusions.

\section{Bosons at filling factor one}

In this section we will analyze the way in which the system of bosons inside a LLL at filling factor one may develop  a paired Pfaffian state. Despite ample evidence from the numerical experiments, on sphere and torus \cite{Canright.Girvin, PhysRevLett.87.120405,PhysRevLett.91.030402,PhysRevA.72.013611,CooperRapidly} that the Pfaffian is the most likely candidate for the description of the ground state, the analytical, field-theoretical understanding is missing. The most rigorous, continuum description that is known (for this system), introduced in Ref. \cite{PASQUIER1998719} , and further developed in Ref. \cite{PhysRevB.58.16262}, in the mean-field approach, does not allow a Pfaffian solution. The description is based on a CF representation, and given that Pfaffian can be interpreted as a state with Cooper pairs of CFs, the inadequacy of the mean-field  approach is puzzling.

In the following Sec. IIA, we will propose and describe the mechanism for pairing, Sec. IIB is devoted to a numerical, exact diagonalization. Section IIA begins with a review of the necessary formalism and a previous approach, and then explains the theory with critical phenomena necessary for pairing.

\subsection{Theory of pairing for bosons at filling factor one}

\subsubsection{Introduction}

The continuum, field-theoretical description \cite{PASQUIER1998719,PhysRevB.58.16262} starts with a two-index fermionic field, $c_{m n}$, in an enlarged space, where indices $ m, n = 1, \dots N_\phi  $ denote the states of a chosen basis in a LL.  $ N_\phi $ denotes the number of flux quanta through the system. The left index, $m$, denotes the state of boson and the right index, $n$, denotes the state of its correlation hole,  of the composite object - composite fermion described by field $c_{m n}$, $c_{m n}^\dagger $:
\begin{equation}
\{c_{m n}, c_{n' m'}^{\dagger}\}= \delta_{n,n'} \delta_{m,m'}.
\end{equation}
The density operators that may be introduced are as follows: physical density,
 \begin{equation}
\rho_{n n'}^{L} = \sum_{m} c_{m n}^{\dagger} c_{n' m},
\label{lLDen}
\end{equation}
i.e. the density of bosons, and artificial, unphysical density,
\begin{equation}
\rho_{m m'}^{R} = \sum_{n} c_{m n}^{\dagger} c_{n m'},
\label{RDen}
\end{equation}
referring to the density of correlation holes. 

Given that there are as many correlation holes as bosons, and that correlation holes have fermionic statistics, the necessary constraint that will reduce the number of the degrees of freedom (of the enlarged space) to the physical ones is
\begin{equation}
 \rho_{n n}^R = 1 . \label{nncon}
\end{equation}
On the other hand, one may introduce the description in the inverse space, 
\begin{equation}
c_{\bs{k}}= (2 \pi)^{\frac{1}{2}} \sum_{m,n}  \langle n\vert\tau_{-\bs{k}}\vert m\rangle   c_{mn},
\end{equation}
with $\tau_{\bs{k}} = \exp\left(i \bs{k} \cdot\bs{R}\right)$, where $\bs{R}$ is a guiding-center coordinate of a single
particle, of charge $q = - e < 0$, in the external magnetic field, $ \bs{B} = - B {\bs{e}_z} $, so that
\begin{equation}
[R_x , R_y ] = - i ,
\end{equation}
where we took $l_B$ (magnetic length) $=1$. Thus the composite object - dipole will have momentum ${\bs k}$, if the two (localized) states of the composite are distance $ {\bs e_z} \times {\bs k} $ apart, because $ {\bs e_z} \times {\bs R} $ has the role of translation operator  in a LL.  The inverse relationship exists and includes an integration over the complete $ {\bs k}-$plane \cite{PhysRevB.58.16262}. 

The density operators in the inverse space are as
follows: the physical density operators,
\begin{equation}
\rho_{\bs{q}}^{L} = \int \frac{d\bs{k}}{(2\pi)^2} c_{\bs{k} - \bs{q}}^\dagger c_{\bs{k}} \exp\left(i \frac{\bs{k} \times \bs{q}}{2}\right), \label{BDen}
\end{equation}
that make a GMP algebra, and the unphysical density operators,
\begin{equation}
\rho_{\bs{q}}^{R} = \int \frac{d\bs{k}}{(2\pi)^2} c_{\bs{k} - \bs{q}}^\dagger c_{\bs{k}} \exp\left(- i \frac{\bs{k} \times \bs{q}}{2}\right) \label{BDenr},
\end{equation}
that make a GMP algebra for the density of particles with opposite charges = correlation holes.

The Hamiltonian of the system is simply
\begin{align}
{\cal H} =\frac{1}{2} \int \frac{d{\bs{q}}}{(2\pi)^2} \;{\tilde V} (|{\bs{q}}|)   \label{RHam}
: \rho^{L}({\bs{q}})
 \rho^{L}(-{\bs{q}}):,
\end{align}
with $ {\tilde V} (|{\bs{q}}|) $ denoting a two-body interaction projected to a LL. It is often assumed that the two-body interaction is the delta (contact) interaction projected to the LLL,
$ {\tilde V} (|{\bs{q}}|) = V_0  \exp\left(-  \frac{|\bs{q}|^2}{2}\right) . $ The Hartree-Fock (HF) mean-field approach leads to a well-behaved electron gas description \cite{PhysRevB.58.16262}, but if we attempt a BCS mean-field decoupling we have to conclude that no $p$-wave instability is possible, no non-trivial solution order parameter exists. See Appendix \ref{appendixA} for details.

A new form of the Hamiltonian was proposed for this system in Ref. \cite{PhysRevB.102.205126}, similar to the one already proposed for the dipole representation of the electron  system at filling factor 1/2 \cite{RevModPhys.75.1101}. It is based on the inclusion of the null operator, $\rho_{\bs{q}}^{R}$, ${\bs q} \neq 0$, that describes the constraint, $ \rho_{n n}^R = 1 $, in the inverse space,  $\rho_{\bs{q}}^{R} = 0$ and defines the physical space in the enlarged space. With this inclusion,  $\rho_{\bs{q}}^{L}$ are substituted with $\rho_{\bs{q}}^{L} -\rho_{\bs{q}}^{R}$, and this represents, in the long-wavelength approximation, a gradient of the dipole polarization \cite{PhysRevB.102.205126}. Next, the normal ordering (present in (\ref{RHam})) in the final form of the Hamiltonian is absent. Thus, the dipole nature of the quasiparticles is more pronounced in this form of the Hamiltonian \cite{PhysRevB.102.205126},
\small
\begin{align}\label{DSHam}
{\cal H}=\frac{1}{2}\int \frac{d{\bs{q}}}{(2\pi)^2}{\tilde V} (|{\bs{q}}|)
 (\rho^{L}({-\bs{q}}) - \rho^{R}({-\bs{q}}))
  (\rho^{L}({\bs{q}}) - \rho^{R}({\bs{q}})),\nn\\
\end{align}
\normalsize
and in a mean-field we can get non-trivial results working on the whole enlarged space.  In Ref. \cite{PhysRevB.102.205126}, the described inclusion of $\rho_{\bs{q}}^{R}$ was justified by the clear description of a single-dipole self-energy (a mass term) that the inclusion brings. But in this way  the demand of the invariance under the  translations in the momentum space (the boost invariance) in this purely interacting system is not taken into account and some other questions seem still unresolved. First, the implied description of a FL-like state, in the HF approximation, is based on the form of Hamiltonian that has an extra symmetry under exchange of $L$ and $R$ degrees of freedom, i.e., two densities, $\rho_{\bs{q}}^{L}$ and  $\rho_{\bs{q}}^{R}$.  Second, the BCS mean-field approach  (details can be found in Appendix \ref{appendixB})  generates two pairing solutions, of opposite angular momenta, $ l = - 1$ and $+1$ (due to the extra symmetry), but these do not represent Pfaffian and a partner (under a time reversal in the CF description), because, due to the explicit dipole-dipole interaction in the form of the Hamiltonian, the order parameter behaves as $ \Delta_{\bs k} \sim |{\bs k}|^2 (k_x - i k_y ),$ and this cannot be a canonical Pfaffian, with a distinct pairing function, $ g({\bs r}) \sim 1/z$, at long distance \cite{PhysRevB.61.10267}.  The paired states (that follow from the Hamiltonian in Eq. \eqref{DSHam}) on the level of the CF description, are characterized by the pairing (two-particle, Cooper) wave functions of a nonanalytic form, at long distance (see Appendix \ref{appendixB} for more detail). This non-analytic behavior is incompatible with the allowed behavior of wave-functions inside a LL. Thus, certainly these solutions are ill-defined and unstable, although have lower energy than the FL-like state in the mean-field approach. 

\subsubsection{Theory of pairing}

Therefore, we would like to revisit the problem of bosons at filling factor one, in particular, consider the ways, that in the enlarged space formalism, in the CF representation, may lead to the Pfaffian state as a true representation of the system and its ground state physics.
The effective, low-energy and long-wavelength physics can be described as physics of dipoles, i.e., CF excitations with dipole moments: the composites must have dipole moment in order to have a finite momentum in the external magnetic field (or the internal Berry curvature field of the band that represents a LL) that acts on them and vice versa, i.e., the finite momentum is followed by the finite dipole moment due to their Fermi statistics. This is most pronounced at the Fermi level when the correlation hole is at the largest distance from the elementary boson in an assumed FL-like ground state.  Thus, they avoid each other in an effective picture, and this may be encoded via effective constraint,
\begin{equation}
\rho^L ({\bs q}) + \rho^R ({\bs q}) = 0, \label{con}
\end{equation}
valid for small $|{\bs q}|$. 

This is certainly a phenomenological description, based on the expected (on physical grounds) dipole description, and, in a way, it collides with the microscopic constraint, (\ref{nncon}),
\begin{equation}
\rho^R ({\bs q}) = 0, \label{qcon}
\end{equation}
unless $ \rho^L ({\bs q}) = 0 $. But, at the same time, it tells us that (a) a compressible (FL-like) state is unlikely, and (b), if proven appropriate, it describes a situation in which $L$ and $R$ densities should be considered on equal footing, as if the physics is invariant under their exchange, though microscopically, in the well-grounded formalism \cite{PASQUIER1998719,PhysRevB.58.16262}, they represent densities of localized objects with different statistics and the constraint $\rho^R ({\bs q}) = 0$. We will come back to this question of the relationship between the effective and microscopic descriptions. In fact, we will argue later that the Hamiltonian \eqref{DSHam} or (after simple rescaling) $ H_{\rm eff}$ \eqref{Hameff}, which commute(s) with the constraint, (\ref{con}), (on the physical space, and it is in this way defined) may describe a deconfined quantum critical point  \cite{senthil2023deconfined,PhysRevB.89.235116} for which fermion-number conservation is broken.

Therefore, an effective Hamiltonian which is appropriate for the description of this low-energy, long-wavelength physics must at least commute with this constraint, (\ref{con}),  on the physical space (defined by the constraint). The most natural form of this Hamiltonian that emphasizes the dipole nature of CFs, is 
\small\begin{align}\label{Hameff}
H_{\rm eff} = \frac{1}{8}\int \frac{d{\bs{q}}}{(2\pi)^2}{\tilde V} (|{\bs{q}}|) 
 (\rho^{L}({-\bs{q}}) - \rho^{R}({-\bs{q}}))
  (\rho^{L}({\bs{q}}) - \rho^{R}({\bs{q}})). \nn \\
\end{align}
\normalsize
The form can lead to a well-behaved  description, in the HF mean-field approach \cite{PhysRevB.102.205126}, and in that respect is better defined than
\begin{equation}   
H_{\rm var} = - \int \frac{d{\bs{q}}}{(2\pi)^2}{\tilde V} (|{\bs{q}}|)   \label{VHam}
 \rho^{L}({-\bs{q}})  \rho^{R}({\bs{q}}),
\end{equation}
which is unstable, ill-defined in the HF approach, with a single-particle energy that decreases with ${\bs k}$. Both Hamiltonians can be reduced to the form in the enlarged space representation, (\ref{RHam}), on the physical space, by applying the constraint. But, if we look for the BCS instabilities, that may correspond to Pfaffian in  numerical experiments, both Hamiltonians lead to unsatisfactory description;  $H_{\rm eff}$ to pairing correlations  at  long distances that do not belong to a LL, and $ H_{\rm var} $ has an unstable (negative-mass) single-particle description.  In the mean-field HF-BCS approach, in the case of $ H_{\rm eff}$, a FL-like state has higher energy than the ill-behaved pairing solutions. This certainly tells us that the description with the emergent (higher) symmetry, the symmetry under exchange of $L$ and $R$ degrees of freedom, cannot be a true low-energy description, because, in the first place, it describes unnatural situation, in which a particle is distanced from its correlation hole. The solutions that break this symmetry will be more favorable.

Indeed, by considering the following  Hamiltonians that break this symmetry, and in a long-wavelength, i.e., dipole representation of the interaction,
\begin{equation}
H^{\rm eff}_{\rm Pf} = \frac{1}{4}\int \frac{d{\bs{q}}}{(2\pi)^2}{\tilde V} (|{\bs{q}}|)   \label{pfHam}
 \rho^{R}({-\bs{q}})
  (\rho^{R}({\bs{q}}) - \rho^{L}({\bs{q}})), 
\end{equation}
and,
\begin{equation}
H^{\rm eff}_{\rm Pf^*}=
\frac{1}{4}\int \frac{d{\bs{q}}}{(2\pi)^2}{\tilde V} (|{\bs{q}}|)   \label{pfcHam}
 \rho^{L}({-\bs{q}}) 
  (\rho^{L}({\bs{q}}) - \rho^{R}({\bs{q}})),
\end{equation}
in the HF-BCS mean-field approach, we can find well-defined pairing solutions, $l = -1$, i.e., the Pfaffian state of CFs, in the case of $ H^{\rm eff}_{\rm Pf}$, and  $l = 1$, in the case of $ H^{\rm eff}_{\rm Pf*}$. Details can be found in Appendix \ref{appendixC}. The  long-wavelength, i.e., dipole representation of the interaction, in the expressions (\ref{pfHam}) and  (\ref{pfcHam}), is necessary, because otherwise single-particle energies will not be well-defined. These Hamiltonians can be considered as spontaneous symmetry breaking forms of the $ H_{\rm eff} $ (adapted to the broken discrete symmetry states).

If the Pfaffian state is realized in the system of bosons in the LLL, at filling factor one, we expect that the effective description in the CF representaion is given in (\ref{pfHam}). To find out to which interaction and Hamiltonian, in the elementary boson representation, this is equivalent, in the long-distance approximation, we will use inverse Chern-Simons (CS) transformations \cite{ZHANG, PhysRevB.47.7312}, which are appropriate for that limit and approximation. They capture the most important transformation, i.e., many-body phase (CS) transformation between CFs and bosons. 
In the long-wavelength approximation we may consider the following form of $H_{\rm Pf}^{\rm eff}$:
\begin{align}\label{appfHam}
H^{\rm eff}_{\rm Pf} \approx\frac{1}{4}\int \frac{d{\bs{q}}}{(2\pi)^2} V_0  \rho^{R}({-\bs{q}})\int \frac{d{\bs{k}}}{(2\pi)^2} (- i {\bs k} \times {\bs q})  c^\dagger_{{\bs k} -{\bs q}} c_{\bs k} .\nn \\
\end{align}
The operator $ \bs{K}=\int \frac{d{\bs{k}}}{(2\pi)^2} \;  {\bs k} \;  c^\dagger_{{\bs k} -{\bs q}} c_{\bs k} $ represents the canonical momentum of CFs, and we need to relate this momentum to the one of elementary boson. We recall \cite{ZHANG}  that, in the first quantization, we have that the mechanical momentum of boson $i$, with charge $q = - e$, ${\bs \Pi}_b = {\bs p}_i + \frac{e}{c} {\bs A} $,  in the presence of the external field, $ {\bs \nabla} \times {\bs A} = - B, B >0$, becomes  ${\bs \Pi}_{\rm CF} = {\bs p}_i + \frac{e}{c}{\bs A} - \frac{e}{c} {\bs a} $, i.e. the mechanical momentum of CF, after the CS transformation, for which,
\begin{align}\label{cs}
{\bs a} ({\bs r}_i )=\sum_{j (\neq i)} \frac{(z_i^* - z_j^* )}{|z_i - z_j |} (- i {\bs \nabla} ) \frac{(z_i - z_j )}{|z_i - z_j|}, 
\end{align}
or
\begin{equation}
{\bs \nabla}_i \times {\bs a} ({\bs x}_i ) = 2 \pi \rho ({\bs x}_i ).
\end{equation}
That is, in our conventions, in which the LLL wave functions are holomorphic (depend, up to the Gaussians, only on $z$ coordinates), the CS field is in the direction along the $z$ axis (opposite with respect to the one of the external field).
Thus, in the bosonic field  functional representation, the effective Hamiltonian becomes,
\begin{align}\label{bosfHam}
{\cal H}^{\rm eff(b)}_{\rm Pf} = \frac{1}{4}\int \frac{d{\bs{q}}}{(2\pi)^2} V_0 \;  
 \rho ({-\bs{q}})
  (- i )\; (\bs{K} + {\bs a}) ({\bs q})  \times {\bs q} ,\nn \\
\end{align}
where $  \rho ({\bs q})$ is the unprojected density of bosons (the same as of CFs).  
If we assume that in the long-wavelength description, the term with $ \bs{K}({\bs q})$, the interaction between density and vorticity i.e. vortex density may be neglected, we are left with a three-body term,
\small
\begin{align}\label{bosfHam2}
{\cal H}^{\rm eff(b)}_{\rm Pf}&  \approx
\frac{1}{4}\int \frac{d{\bs{q}}}{(2\pi)^2} V_0   
 \rho ({-\bs{q}}) \int \frac{d{\bs{k}}}{(2\pi)^2}
  (- i )    \rho ({-\bs{k}}) 
    ({\bs a} ({\bs k} + {\bs q})  \times {\bs q})\nn \\
&= \frac{ V_0 }{4}\int \frac{d{\bs{q}}}{(2\pi)^2}\; \int \frac{d{\bs{k}}}{(2\pi)^2}\; \int d{\bs x}_1  \; \int d{\bs x}_2 \; \int d{\bs x}_3  \nn \\  
&   \;\;\;\;\;   \times \exp\left(- i \bs{q} \cdot \bs{x}_1 \right)    \exp\left(- i \bs{k} \cdot \bs{x}_2 \right)    \exp\left( i  (\bs{q}+\bs{k}) \cdot \bs{x}_3 \right)  \nn \\
&   \;\;\;\;\;   \times \rho (\bs{x}_1) \; \rho (\bs{x}_2)  \; {\bs \nabla} \times {\bs a} (\bs{x}_3 )\nn \\
& = \frac{2 \pi}{4} V_0  \int d{\bs x}_1  \; \int d{\bs x}_2 \; \int d{\bs x}_3  \; \delta^2 ({\bs x}_1 - {\bs x}_3 ) \; \;  \nn \\
& \;\;\;\;\;  \times \delta^2 ({\bs x}_2 - {\bs x}_3 )\rho (\bs{x}_1) \; \rho (\bs{x}_2)  \; \rho(\bs{x}_3). 
\end{align}
\normalsize
Thus we find that the effective interaction (in the long-distance approximation) for bosons is the three-body interaction, a product of two delta functions,
\begin{equation}
H_b^{\rm eff} = \sum_{<ijk>}  \delta^2 ({\bs x}_i - {\bs x}_j ) \; \; \delta^2 ({\bs x}_i - {\bs x}_k ),
\end{equation}
i.e. the one that is considered as a model interaction for Pfaffian. This also strengthens our belief that $H^{\rm eff}_{\rm Pf} $ represents the effective Hamiltonian based on the two-body interaction between bosons, that due to the constraints (the topology) of the space they live in, they may end up in the state that is the ground state of $H^{\rm eff}_{\rm Pf} $. 

We can apply the same reasoning that lead us to the derivation of the three -body interaction among bosons in the case of Pfaffian and $H^{\rm eff}_{\rm Pf} $, to the case of the Hamiltonian $H^{\rm eff}_{\rm Pf*} $. We can immediately conclude that in this case the three-body interaction  will be attractive and lead to some kind of Bose-Einstein condensation in real space. Likely this represents an unstable state or an ordered state with a broken translational symmetry \cite{PhysRevLett.87.120405}, which gives way to a Pfaffian phase, as seen in numerical experiments.  The phase described by $H^{\rm eff}_{\rm Pf*} $ may correspond to the one recently discussed in Ref. \cite{musser2023observable}.

But, on the other hand, the phase of bosons that interacts via the contact interaction, and has some features, like the ground state of Pfaffian phase, is not only represented by the ground state but also by excited states, in the presence of the two-body interaction. The excited states may not have topological features (as found in the case of three-body interaction) as discussed in Ref. \cite{PhysRevB.92.075116}, and, also, this may be a reason for a weak entanglement features for the system on lattice as described in Ref. \cite{PhysRevB.86.165314}. Pfaffian physics may be plagued by the time-reversed partner in the CF representation. Thus, a more elaborate analysis is needed.

We may wonder if we can modify the continuum, microscopic description of Ref. \cite{PhysRevB.58.16262} by simply adding terms with the exact constraint, $ \rho^R ({\bs q}) = 0, {\bs q} \neq 0 $, to the basic Hamiltonian, (\ref{RHam}), and thus adapting the description to a mean-field procedure, that will lead to Pfaffian. Thus, we may consider, in the LLL,
\small
\begin{align}
H^{\rm mod} ={\cal  H} + \int \frac{d{\bs{k}}}{(2\pi)^2} C \exp(- \frac{q^2}{2})  \rho^{R}({-\bs{q}})
  (\rho^{R}({\bs{q}}) - \rho^{L}({\bs{q}})). \label{modHam}
\end{align}
\normalsize
But soon we will realize (when we consider the angle integration in a self-consistent equation for order parameter) that $C $ must be $C > 2$, i.e., effectively, a strong enough three-body interaction is needed to stabilize Pfaffian.

\begin{figure}[h!]
	\centering
	\includegraphics[width=0.95\linewidth]{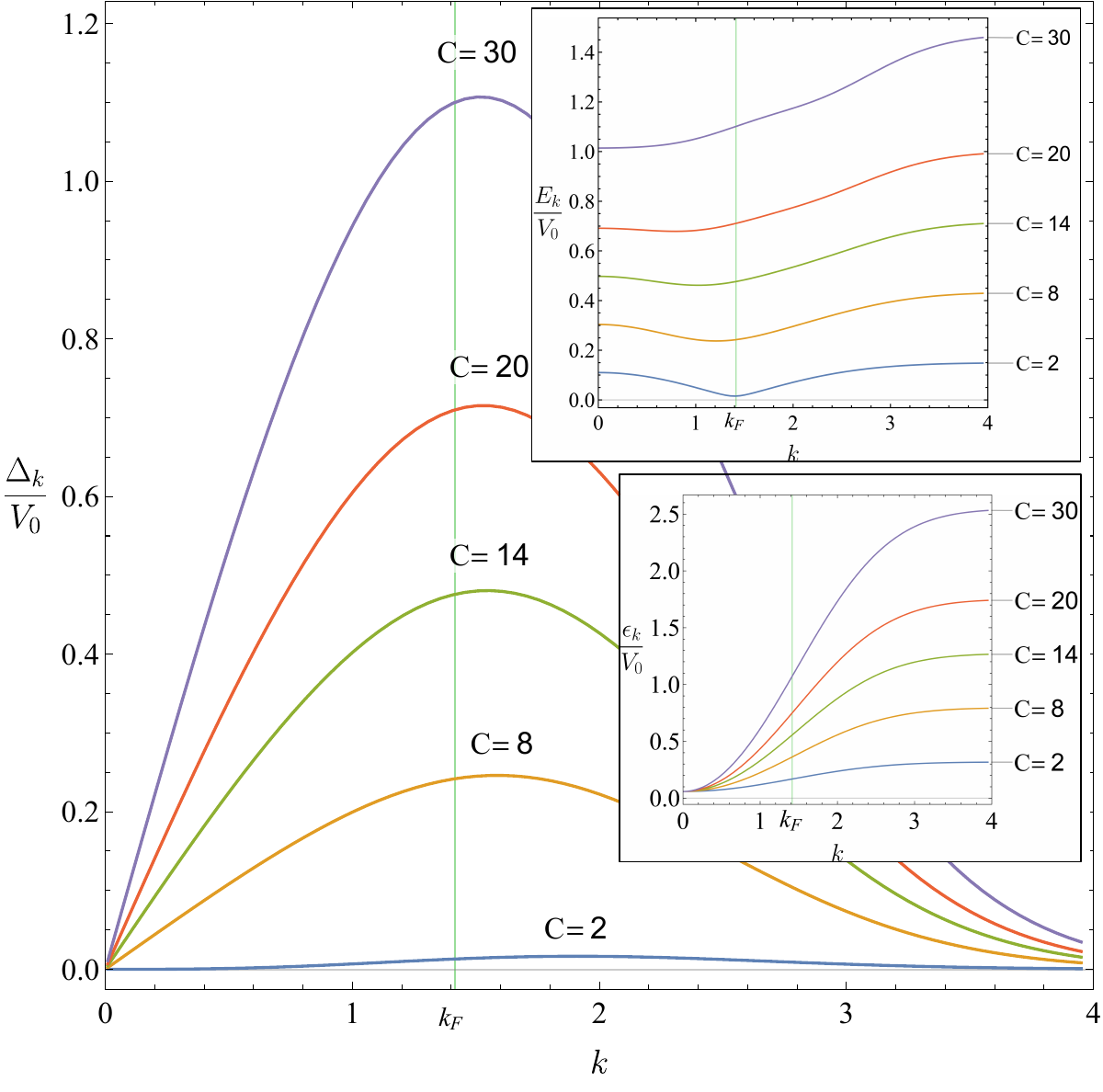}
	\caption{{Graphic of the order parameter $\Delta_{\bs{k}}$, the Bogoliubov energies $E_{\bs{k}}$, and the dispersion $\epsilon_{\bs{k}}$ as functions
of $|\textbf{k}|$ for few values of $C$.}}	\label{FigOrderParBogoliubovDisBoz}
\end{figure}

In Fig.\ref{FigOrderParBogoliubovDisBoz}  plotted are the absolute value of the mean field parameter $\Delta_{\bs{k}}$ and Bogoliubov energies $ E_{\bs{k}}$ as functions of $|\bs{k}|$ for few values of $C$. We can recognize the weak to strong coupling evolution as $C$ value is increased.

 We can recover the well-known description of Pfaffian in the  large-$C$  limit.  The details of the mean-field solutions can be found in Appendix \ref{appendixC}. We can conclude that the Pfaffian instability in its ideal form is  associated with a critical behavior with Pfaffian (features) being the strongest in the neighborhood of a  critical point described by $ H_{\rm eff} $. The $ H_{\rm eff} $  represents a description of the critical point, between the $ l = 1$ and $-1$ pairing, at which there is a release (deconfinement) of neutral fermions - Majoranas, the fermionic number is not a conserved quantity,  and the constraint $\rho^R ({\bs q}) = 0$ is valid only on average. Indeed, if we follow the usual (vortex problem) set-up for Majorana solution of Ref. \cite{PhysRevB.61.10267}, due to $ \Delta_{\bs k} \sim |{\bs k}|^2 (k_x - i k_y ),$ when $|{\bs k}| \rightarrow 0$ (and thus $ \Delta_{\bs k}$ can be neglected with respect to $ \epsilon ({\bs k}) \sim |{\bs k}|^2 $, single-particle energy), the zero-energy solution will be delocalized (instead of exponential decay we will have oscillating behavior). Detailed derivation of this behavior at the critical point as well a detailed description of the critical pairing solutions  described by $H_{\rm eff}$ in Eq. \eqref{Hameff} can be found in Appendix \ref{appendixB}.
 
In our effective description we do not know the value of $C$ that corresponds to the system with the contact interaction; the appropriate weight as given  by $C$ can be a value greater than $2$. The numerical experiments  \cite{PhysRevA.77.063602,Regnault_2006,PhysRevA.77.043608} that explored the influence of the second relevant Haldane pseudopotential, $V_2$, found that beyond $\alpha = V_2 / V_0 \approx 0.3 $, a phase with stripes exists. The critical point is the point of collapse of a well-developed Pfaffian phase into the phase with stripes. The value of the overlap of the ground state with the Pfaffian state is at maximum at or near the critical point, in accordance with the expectation based on our description of paired states as the states that accompany  and originate  from the critical behavior.

We may also notice that $ H^{\rm mod}$, with an explicit single-particle term, violates the boost invariance necessary for the  FL state description \cite{PhysRevB.107.155132}. The only modification, without this violation, is $ {\cal H} \rightarrow {\cal H} -  \int \frac{d{\bs{k}}}{(2\pi)^2} {\tilde V} (|{\bs{q}}|)   \rho^{R}({-\bs{q}})  \rho^{R}({\bs{q}}) $, which leads to pairing in the $l = 1$ channel. But without a single-particle term  a pairing is ill-defined or represents a gapless phase.  To conclude, a single-particle term is expected in a gapped paired state, and thus appears in the effective Hamiltonians $H^{\rm eff}_{\rm Pf}$ and  $H^{\rm eff}_{\rm Pf*}$, which are based on the spontaneous symmetry breaking of the parent Hamiltonian, $H^{\rm eff}$, which also has a single-particle term, as a consequence of the effective dipole physics and the emergent symmetry.   The presence of this term is a problem that we have to solve if we want to have a well-defined FL state description \cite{PhysRevB.107.155132}, with the boost invariance ($K$), i.e., a FL state inside a LL.

We find a further support for considering $H^{\rm eff}$ as a critical, unstable description simply by finding a way to a well-defined Fermi-liquid description \cite{PhysRevB.58.16262}  (regularizing  $H^{\rm eff}$ by imposing the boost invariance) by eliminating the mass (quadratic) term in the effective (mean-field) CF dispersion as in \cite{PhysRevB.107.155132}.  Concretely, we consider $H^{\rm eff} + \int \frac{d{\bs{k}}}{(2\pi)^2} {\tilde C} \exp(- \frac{q^2}{2})  (\rho^{R}({-\bs{q}}) + \rho^{L}({-\bs{q}}))
  (\rho^{R}({\bs{q}}) + \rho^{L}({\bs{q}}))$ and choose ${\tilde C}$ to eliminate  mass term. The result is simply $ \int \frac{d{\bs{k}}}{(2\pi)^2}  \frac{V_0}{2}  \exp(- \frac{q^2}{2})  (\rho^{L}({-\bs{q}}) 
  (\rho^{L}({\bs{q}}) + \rho^{R}({\bs{q}}) \rho^{R}({-\bs{q}}))/2$ , a symmetrized version of the Hamiltonian (\ref{RHam}). (The normal ordering can be introduced because it implies a constant, trivial shift.) This Hamiltonian is of the form of the Read's description - in the mean-field approach they do not differ, but we have to bear in mind that the Hamiltonian and description (with respect to the Read's description) is effective, involving only low-energy, long-distance degrees of freedom. Therefore, depending on its interaction, the system of bosons will stabilize in the Fermi-liquid(-like) state, or one of the paired states of CFs. We may expect that due to the description encoded in  $ H^{\rm mod}$, with ${\cal  H}$ as the Hamiltonian of a regularized FL-like state, the Pfaffian state as a BCS instability may be a dominant phase in the bosonic system with the contact interaction.  Below, we will use the exact diagonalization on sphere, to show, on the level of total energies, that in the extrapolated thermodynamic limit, we can detect Pfaffian features.

\subsection{Numerical experiments}

The  numerical experiments, both, on torus and sphere point to the Pfaffian state as the effective description of the ground state for the system of bosons in the LLL that interact via the contact interaction $V_0e^{-\frac{|\bs{q}|^2}{2}}$ \cite{PhysRevLett.87.120405,PhysRevLett.91.030402,CooperRapidly}. Though on sphere, the shift clearly may have the role of the small perturbation necessary to stabilize Pfaffian, on the torus there is no obvious symmetry breaking parameter. As we already discussed , we do not expect the topological behavior from the state of CFs with $l = 1$ pairing, and the three (not so well) degenerate states on torus \cite{CooperRapidly} may describe a stable Pfaffian ground state and physics in the system of bosons. The paired states with $l = 1$ and $l = -1$, on the level of bosonic physics, certainly differ, due to the necessary projection to the LLL and the absence of the symmetry on the microscopic level, and this leads to the prevalent Pfaffian physics. 

\begin{figure}[h!]
	\centering
	\includegraphics[width=0.9\linewidth]{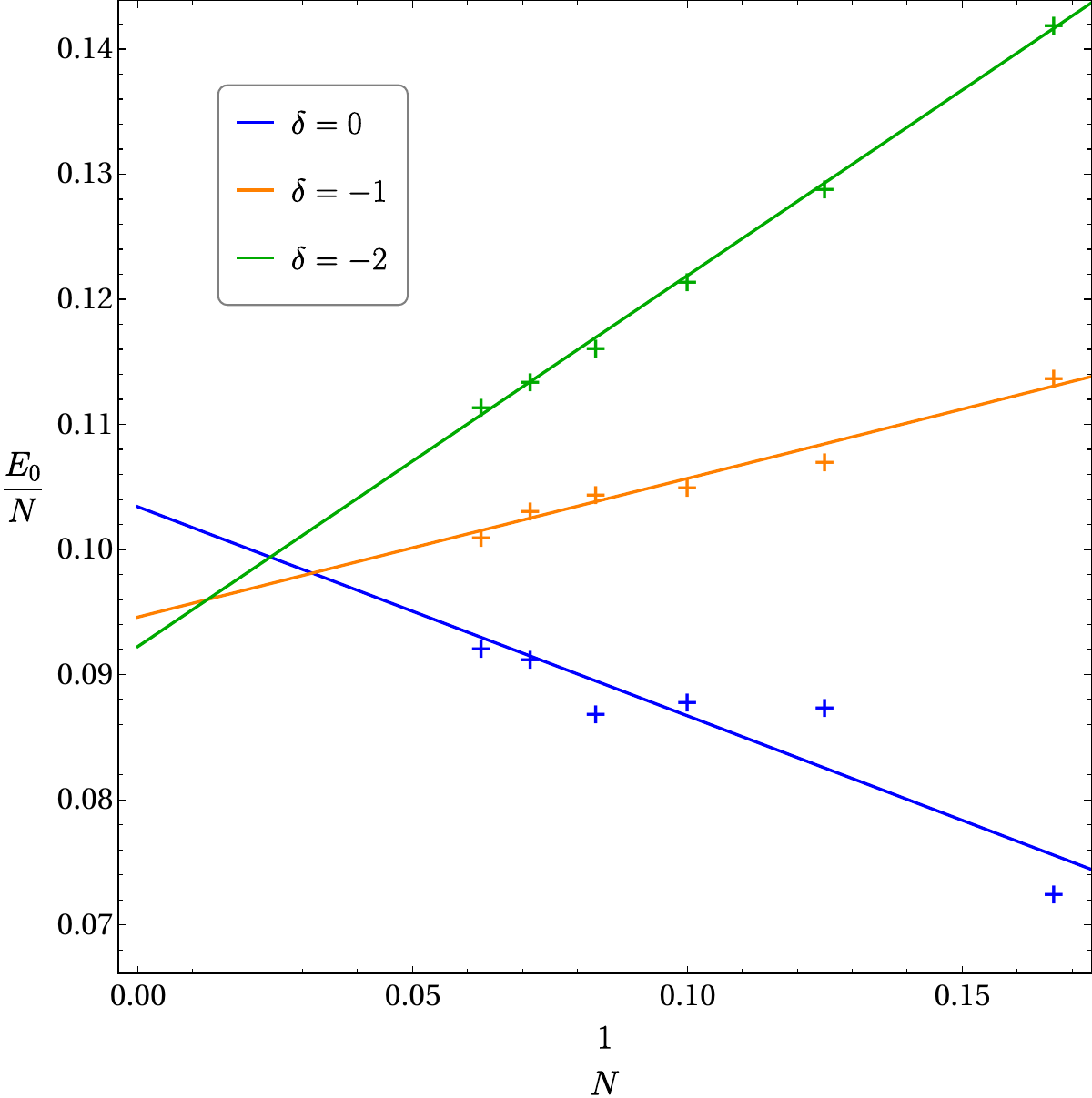}
	\caption{{Graphic of extrapolation of the ground energies $E_{0}$ per particle to the thermodynamic limit, for shifts $\delta=0,-1,-2$. The energy scale is $V_0/l_B^2$ and standard intercept errors of linear fits are 0.004, 0.002 and 0.0008 for shifts $\delta=0,-1,-2$ respectively. The ground state energies are obtained using DiagHam libraries  [\href{http://www.nick-ux.org/diagham/}{http://www.nick-ux.org/diagham/}].}}	\label{FigEkstrapolation}
\end{figure}

To examine  the nature of the ground states, we calculated and compared the ground state energies of the system(s) on the sphere, at different values of shift : (a) shift $\delta= - 2$, characteristic for Pfaffian, (b) shift $= - 1$, which corresponds to the FL state, and (c) shift $= 0$ that is characteristic for the state with $l = 1$ pairing of CFs.  The ground state energies at different shifts and for different system sizes are plotted in Fig. \ref{FigEkstrapolation}. 
The Pfaffian is a well-established phase in the thermodynamic limit; the ground state is at shift $= -2$, and  the states at shift $= - 1$, and shift $= 0$  represent one and two quasihole excitations, respectively, which energies describe the increase in energy with each quantum of flux that is introduced in the system. 

\section{Electrons at filling factor 1/2}

\subsection{ Introduction}

In this section we will consider paired states of half-filled LL of electrons in the dipole representation. We can introduce the dipole representation for this system by applying the same reasoning as in the bosonic case that lead us to the Hamiltonian $H^{\rm eff}$ (\ref{Hameff}). More details and explanations can be found in Appendix \ref{appendixD} and in Ref. \cite{PhysRevB.107.155132}.

The form of the Hamiltonian is the same as in the bosonic case, $H^{\rm eff}$,  but here the symmetry that is encoded in $H^{\rm eff}$, i.e. the symmetry between $L$ and $R$ degrees of freedom, in this case coincides with the PH symmetry (exchange between particles and holes) that in this system exists in the microscopic description, on the level of electrons; because $R$ degrees of freedom represent  correlation holes, the constraint, 
\begin{equation}
\rho^L ({\bs q}) + \rho^R ({\bs q}) = 0, \label{econ}
\end{equation}
effectively (in the long-distance $|{\bs q}| \sim 0 $) equates the density of correlation holes with the density of real holes (fermions) and, in this sense, via invariance under exchange of $R$ and $L$ densities the PH symmetry is present in $H^{\rm eff}$.

Thus, the breaking of the symmetry under exchange of the densities of $L$ and $R$ degrees of freedom in $H^{\rm eff}$ can be identified with the breaking of the PH symmetry, which is exactly the particular symmetry and breaking that is associated with the Pfaffian \cite{PhysRevLett.99.236807, PhysRevLett.99.236806}. Thus a two-body microscopic  interaction may be all that is needed to realize Pfaffian as emphasized in \cite{PhysRevB.99.045126}. The associated symmetry (with its breaking) is not an emergent, effective but present at this microscopic level (dipole physics is more pronounced in this system) and may lead to a stable Pfaffian phase. 

We may consider what is the interaction among electrons that corresponds to a two-body interaction between dipoles in $H^{\rm eff}_{\rm Pf}$. We will consider $ V(|{\bs q}|) = -|{\bs q}|^2$, $ {\tilde V}( |{\bs q}|) =  V(|{\bs q}|) \exp(-  |{\bs q}|^2 /2) $, i.e.,  a simple, analytic, short-range, repulsive interaction, and apply similar arguments as in the bosonic case. In the long-wavelength expansion, a first, three-body contribution (that affects electrons) is of the following form:
\begin{equation}\label{3bodyelec}
H_e^{\rm eff} = \sum_{<ijk>}  {\bs \nabla}^2_i  \delta^2 ({\bs x}_i - {\bs x}_j ) \; \; \delta^2 ({\bs x}_i - {\bs x}_k ),
\end{equation}
which is the usual expression for the model interaction in the case of a Pfaffian.  Detailed derivation of this  three-body interaction can be found in Appendix \ref{appendixE}. If we assume  $ V(|{\bs q}|) = -|{\bs q}|^6$, using the similar steps as in the Appendix \ref{appendixE} we recover the model interaction proposed in Ref. \cite{PhysRevLett.84.4685}.

On the other hand, as in the bosonic case, the state that corresponds to $H^{\rm eff}_{\rm Pf*}$, i.e. its solution in the mean-field approach to $l=1$ pairing, may not correspond to a stable phase. This should  correspond to the absence (criticality) of so-called PH Pfaffian state in an isolated LL, as conjectured in Ref. \cite{PhysRevB.95.235304} , and later confirmed in numerical experiments \cite{PhysRevB.98.035127,PhysRevB.98.081107,PhysRevB.102.195153}.

If we switch the sign of charge of the elementary particle - electron, we will describe the physics from the point of view of holes in that case,  and we may reach a stable anti-Pfaffian case, with the $l = 1$ instability, in the dipole representation.  (We will have the CS transformation in the opposite sense, and the sign of the three-body interaction will also flip. Thus, we have to perform an additional  (PH) transformation also in the CS transformation, to be able to identify the description by $H^{\rm eff}_{\rm Pf*}$ as the one of anti-Pfaffian.) 

Because in this case the symmetry is not just emergent, effective, but exists on the microscopic level, we can expect that the dipole representation and $H^{\rm eff}$ is more appropriate and valid in a larger energy range in the electron, half-filled systems, with two-body interaction than in the bosonic system, and thus implied solutions may be less prone to changes, more stable and topological in nature, which is certainly the case according to numerical experiments \cite{PhysRevB.99.045126,PhysRevB.92.075116}.

\begin{figure*}[!tp]
\hfill%
     \subfloat[Lowest LL ($n = 0$).\label{FigCutoffa}]{ \includegraphics[width=0.44\linewidth]{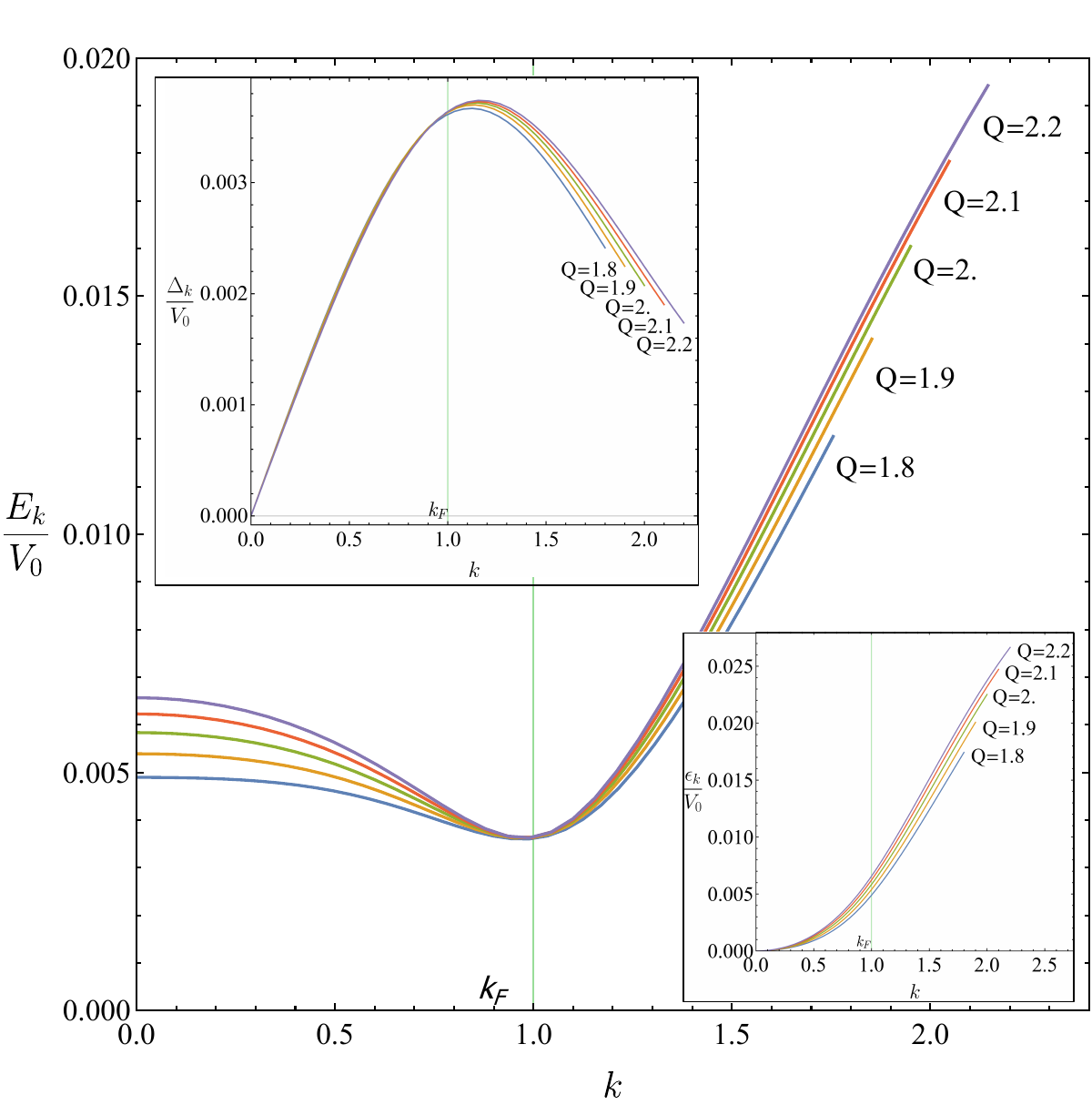}}
     \hfill%
      \subfloat[The second LL ($n = 1$).\label{FigCutoffb}]{ \includegraphics[width=0.44\linewidth]{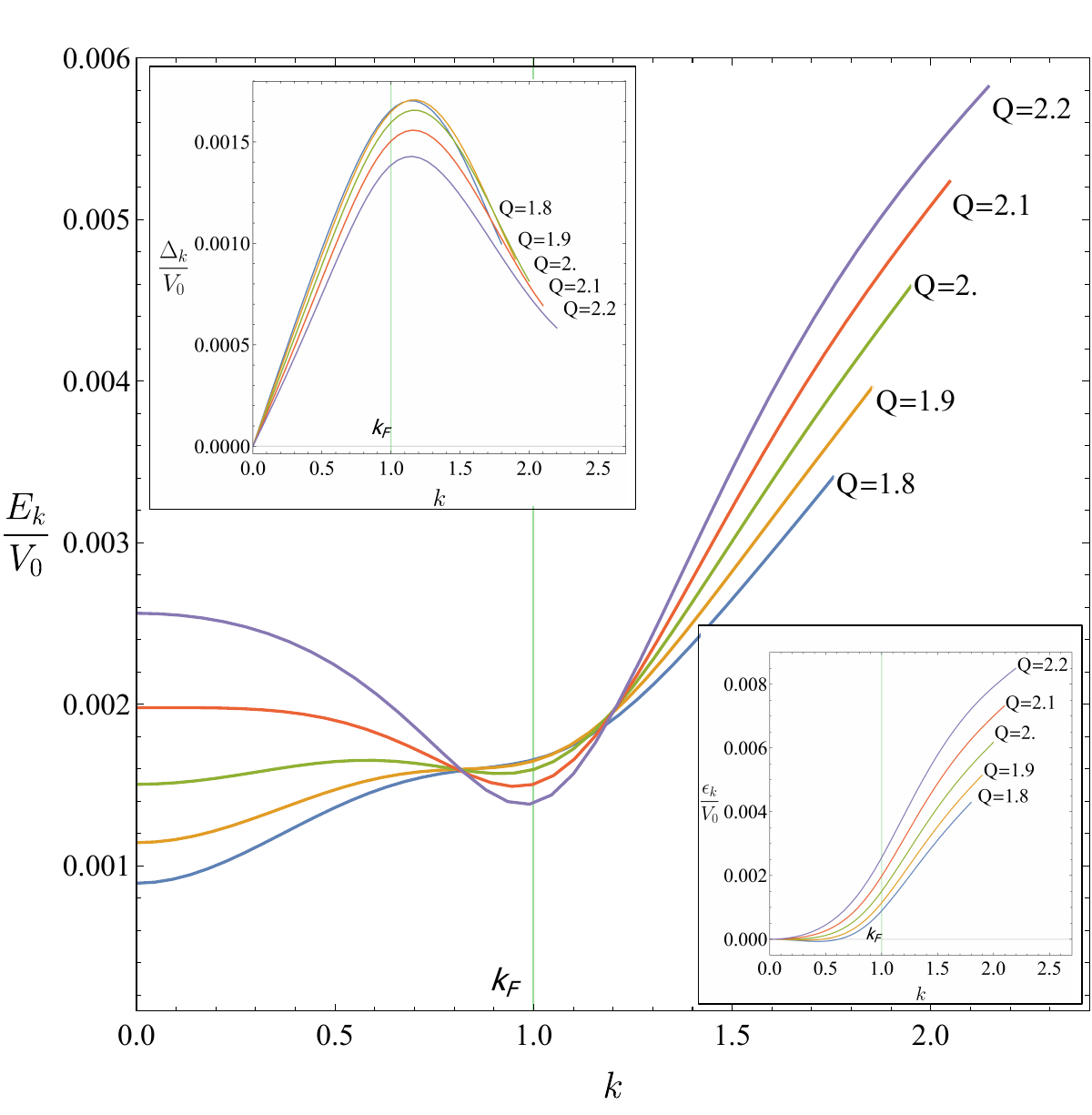}}
      \hfill%
      \hfill%
   \caption{{Bogoliubov quasiparticle dispersions (a) in the LLL ($n = 0$) and (b) in the second LL ($n = 1$), for an interval of values of cut-off $Q$, $Q \in [1.8, 2.2] k_F $ , where $k_F = 1/l_B$. In top-left insertions are plotted the order parameters, and in bottom-right are dispersions for corresponding LL and for same interval of values of cut-off $Q$. The energy scale is set with $V_0=4\pi e^2/l_B$.}}	
    \label{FigCutoff}
\end{figure*}

Nevertheless, as emphasized in the Introduction, the critical behavior (i.e., underlying interaction) leads in some cases to a FLL state (instead of paired state(s)). In the scope of the BCS mean-field approach we would expect that the pairing solutions dominate (if possible and present in the systems at hand), but here we have, what we may call, a critical FLL state, which can stabilize as paired states, but also a regularized FLL state, a distinct state that does not represent a parent state for possible pairing instabilities. The regularized FLL state, due to the absence of the mass (term), does not support gapped paired states. It is in a competition with the paired states, and depending on the interaction and the LL specification, it is possible that the regularized FLL state describes the state of the system.

\subsection{Pairing of electrons at half-filling of a Landau level}

Thus, also in the electron case, we want to find pairing solutions of the Hamiltonian for CFs in (\ref{pfHam}), but in the case of the Coulomb interaction in the lowest LL ($n = 0$), and the second LL ($n = 1$).  In the bosonic case our conclusions on pairing do not change if we introduce a ``cut-off ", a bound on the value of the momentum of the order of $ k_F = \sqrt{2}/l_B $. On the other hand, in the case of electrons at half-filling, inside a LL, the dependence on a cut-off in the momentum integration turns out to be more  subtle. We must bear in mind that in this case we deal with an effective theory and description at the Fermi level with the Coulomb interaction, and physics thus may depend more strongly on the value of the cut-off. But there is also a more fundamental reason for the inclusion of a natural, intrinsic length, i.e., a cut-off in the momentum integration, in a mean-field approach.
The problem of particles (bosons or electrons) in an isolated LL bear similarity to a crystal problem, because of a discretization of space, but not in the way of ``shape"; the discretization is in the way of ``area". Thus an intrinsic length exists, and it can be identified with $l_B$. In contrast to the numerical experiments, when one fixes a shape (i.e. gauge) of the system, and sets up an exact diagonalization of a finite system, in the thermodynamic limit, in the absence of a fixed shape, it is natural to choose a rotationally symmetric domain of allowed values of momenta.

If, we do not use any cut-off in integration, a basic picture that emerges is in agreement with the phenomenology: a regularized FL- like state is dominant in the LLL, while the p-wave pairing of CFs prevails in the second LL. What is unsatisfactory is that the pairing state in the second LL is in the weak-coupling regime, which is not the case according to numerical experiments in the second LL. The most recent numerical studies, which generalize and optimize p-wave weak-pairing correlations in Ansatzes for the ground state function, give different estimates for ratio of the kinetic and pairing energies on torus \cite{Sharma_2021} and sphere \cite{henderson_moller_simon_2023} for the Coulomb type of potential, but in the favor of the strong-coupled case. Thus we are motivated to study a general case in the presence of a cut-off. The role of the cut-off here is similar to the one of the variational parameter – the Debye length used in Refs. \cite{Sharma_2021, Sharma_2024}.

In the following, in the first part, we will introduce a general cut-off, the same for all relevant integrations, in order to understand basic dependence on its value. This analysis will lead to conclusions on the nature of instabilities in the half-filled LLs. In the second part, we will introduce physically motivated values on the bounds on the momentum integration, i.e., cut-off-s in the integration. The conclusions will be the same as in the first part with a general, universal cut-off.

\subsubsection{An approach with a universal ``cut-off'' }\label{An approach with a universal "cut-off"}

The CF momenta can be identified with the momenta of particles and holes (constituents of the CF dipole). Thus, the volume of the momenta that CFs may have, have been bounded by the radius  $ \sqrt{2}/l_B $, which corresponds to the number of single-particle states in a LL. But to include exchanges, i.e., interaction among CFs of different momenta, we will consider a possibility that the exchange momentum is approximately twice as large as the natural bound, $ \sqrt{2}/l_B $. In fact we will explore possibilities for pairing solutions, allowing a general cut-off (which may come (as a necessary bound) certainly as a consequence of our effective description around the Fermi level in the case of electrons).

In Fig. \ref{FigCutoff} we plotted Bogoliubov quasiparticle dispersions (as an outcome of the BCS-HF mean-field treatment of $H^{\rm eff}_{\rm Pf} $  with Coulomb interaction) in the LLL ($n = 0$) and in the second LL ($n = 1$), for an interval of values of cut-off $Q$, $Q \in [1.8, 2.2] k_F $ , where $k_F = 1/l_B$.  The insets show the solutions that we get for the order parameter and CF dispersions. We may notice that the description in the second LL may be characterized for certain values of $Q$, $Q \in [1.8, 2.0] k_F $ , as the one of a strong coupling regime (expected to characterize the Pfaffian state in the second LL). What is puzzling is that, also in the LLL, pairing solutions are present, and this may lead us to believe that in the thermodynamic limit the pairing is present, in the LLL, though this is contrary to numerical experiments that find that a FLL state is the most stable state. But the final verdict (for the absence or presence of pairing) comes from the comparison of the total energies of a pairing solution  and a regularized FLL state, in which description of the demand for boost invariance is incorporated. As we will show below, indeed, our description supports a Pfaffian pairing in a strong coupling regime, in the second LL, and a regularized FLL state in the LLL.

But as an introduction to that analysis we must justify  a value or at least an interval of $Q$ (``cut-off") that may correspond to the problem at hand. To conform to the phenomenology, we analyzed the solutions for the interval of $Q$ values in Fig. \ref{FigCutoff} that exemplify the strong coupling regime of pairing in the second LL, and still have well behaved dispersions (as functions of ${\bf k})$ for the CFs. (Weaker $Q$ will lead to a deep minimum in the CF dispersion for ${\bf k} \neq 0$. Here we restrict solutions to those that are well defined (as a Fermi gas) even for $\Delta=0$, though this limit does not represent the FL state of the system).
As depicted in Fig. \ref{FigCutoffb}, strong-coupling is evident for a cut-off $Q$ smaller than $2$, while the dispersion exhibits relatively good behavior for $Q>1.8$. The term ``relatively'' is used because for $Q$ values between $1.8$ and $1.9$, there is a slight deviation, a dip, yet in our effective theory it remains far from $k_F$, and the dispersion increases monotonically at a sufficiently large distance from $k_F$. Therefore, we focus on the range of $Q\in[1.8,2]$. 

\begin{figure}[t!]
	\centering
	\includegraphics[width=0.8\linewidth]{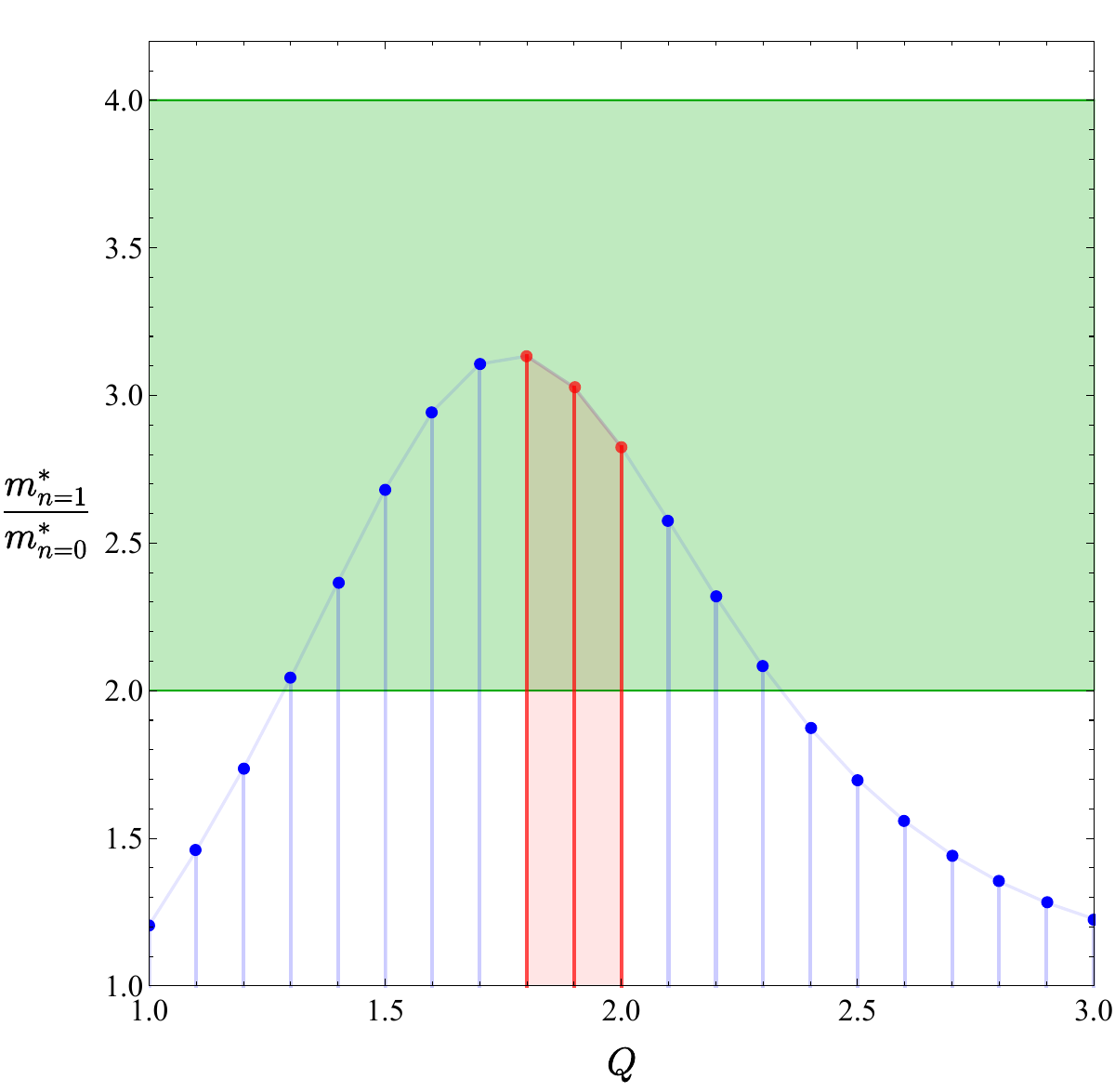}
	\caption{Effective masses rations for different cut-off $Q$.  Green shaded area (ranging from 2 to 4) depicts all the values of rations which are in agreement with the estimates given in Ref. \cite{Petrescu2023}. Red dotes represent values obtained for focused range of cut-offs, $Q\in[1.8,2]$.  }	\label{FigEffMasses}
\end{figure}

Furthermore, if we calculate the effective masses at the Fermi level, of the unregularized FLL state (which we assume will become relevant with LL mixing), in the LLL and second LL, their ratio $m^*_{n=1}/m^*_{n=0}$ is within the interval experimentally estimated in a recent paper \cite{Petrescu2023}, if $Q\in[1.8,2]$  ( Fig. \ref{FigEffMasses}).

Also, increasing $Q$ will lead to weak coupling regimes of pairing, while decreasing $Q$ to strong coupling regimes in both systems.

Now that we have established the relevant interval of $Q$, we can move on to the comparison of the total energies of a pairing solution,

\begin{equation}
E_{\textsc{tot}}^{\rm Pf}=\sum_{k=0}^Q\left(\frac{\xi_{\bs{k}}-E_{\bs{k}}}{2}+\frac{|\Delta_{\bs{k}}|^2}{4E_{\bs{k}}}\right),
\end{equation}
and a regularized FLL state,

\begin{equation}
E_{\textsc{tot}}^{\textsc{fll}}=\sum_{k=0}^{k_F}\left(\epsilon_1(k)+\frac{1}{2}\epsilon_{\textsc{hf}}(k)\right).
\end{equation}

\begin{figure}[!h]
\subfloat[Lowest LL ($n = 0$).\label{FigTotEa}]{ \includegraphics[width=0.8\linewidth]{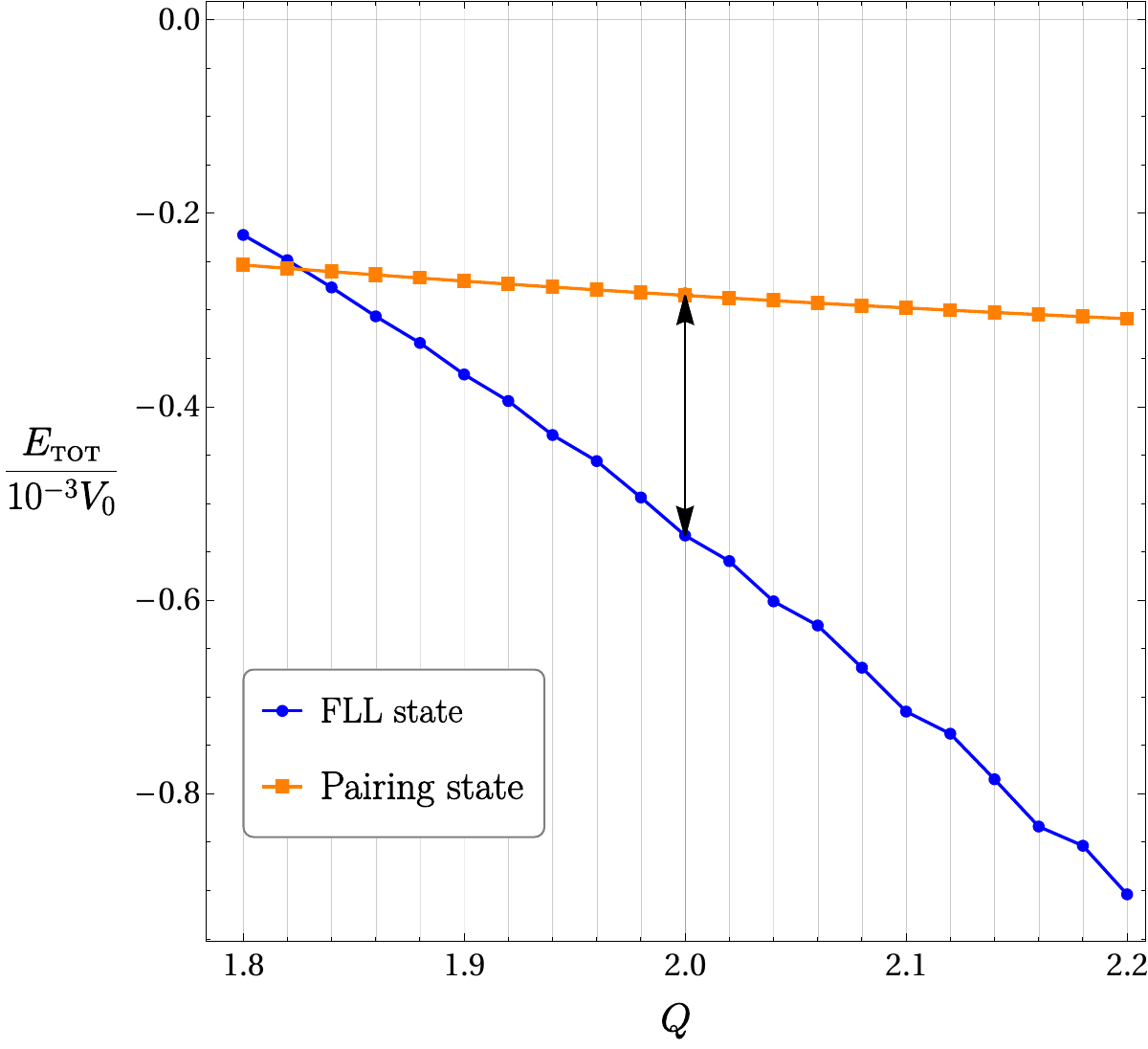}}\\
    \subfloat[The second LL ($n = 1$).\label{FigTotEb}]{ \includegraphics[width=0.8\linewidth]{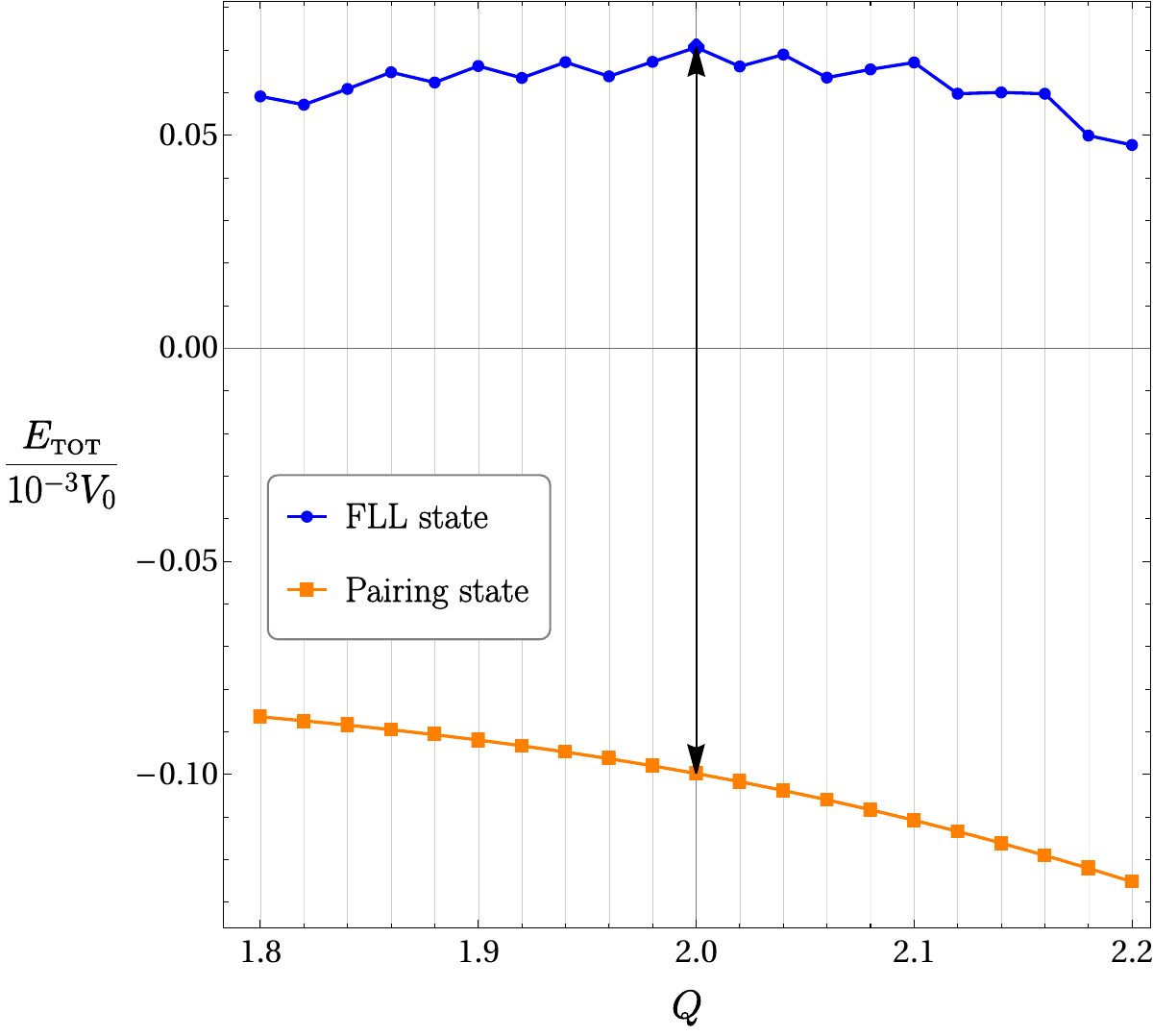}}
   \caption{{ Total energies of a pairing solution  and a regularized FLL state (a) in the LLL ($n = 0$) and (b) in the second LL ($n = 1$), for an interval of values of cut-off $Q$, $Q \in [1.8, 2.2] k_F $, where $V_0=4\pi e^2/l_B$ sets the energy scale.}}	
    \label{FigTotE}
\end{figure}

While the mean-field set-up and ensuing expression for the total energy for the pairing \textit{Ansatz} is straightforwardly established, we need to pay more attention when we set up the HF approach for the regularized FLL state. As detailed in Appendix \ref{appendixD} the interaction generated HF contribution to the overall mass is an artifact of the approach and we need to find ways inside the HF approach (with constraints) to eliminate it. While we can find  Fermi liquid parameters in a description based on the expectation of the quantum Boltzmann equation (see Appendix \ref{appendixD} and Ref. \cite{PhysRevB.107.155132}), to calculate the total energy we stay inside the boundaries of the HF approach, and should find a way to eliminate the mass term (in the boost-invariant system that we describe, including the mass term, i.e., single-particle term inherent to the Hamiltonian in \eqref{Hameff}). Although choosing an appropriate counter term (which is zero in the physical space, but is effective in the mean-field approach), has a limited effect in the case of the HF-generated mass term (see Appendix \ref{appendixD}), we applied  this procedure while calculating the total energy. The overall trend is that, while demanding the boost invariance in this way, the total energy of the regularized FLL state decreases, and indeed, in the case of $n=0$ LL it is of lower value than that of paired state (see Fig. \ref{FigTotE}).

\subsubsection{Physics behind cut-off-s }

A close inspection of the integrals involved in the momentum integration for the total energies leads to the conclusion that the nature of pairing instability as well the likelihood of the pairing depends on the value of the momentum cut-off in the integration for the ``self-energy'' of the dipole \cite{PhysRevB.102.205126}, i.e., the kinetic term with mass, which is necessary for a pairing solution; the smaller the cut-off, the smaller the kinetic term and, therefore, the pairing state is more strongly coupled and more likely to occur. The cut-off in this integration we will denote with $\Lambda_q$. We may connect its value with the inverse of the screening length of the plasma that can be associated with the Pfaffian system. Namely, this length can describe the spatial extent of the well-localized correlation hole (and, like in the classical electrodynamics, the necessary extent of the point-like charges, in order to make the self-energy of the dipole meaningful (not diverging)). This length (that delineates the short-range physics) should be 
a characteristic length for our effective description, though its exact value we can only estimate to be of the order of $l_{B}$; the weakly-coupled plasma approach of Ref. \cite{PhysRevB.83.075303} gives 
an estimate of  $1.2l_{B}$. In the Laughlin strongly-coupled plasma we expect $\sqrt{2}l_{B}$  and thus $\Lambda_q=1/(\sqrt{2}l_{B})$ \cite{prange1990quantum}.

\begin{figure}[h!]
     \subfloat[\label{cut-off-sa}Pairing solution with $\Lambda_q=1/\sqrt{2}$]{ \includegraphics[width=0.75\linewidth]{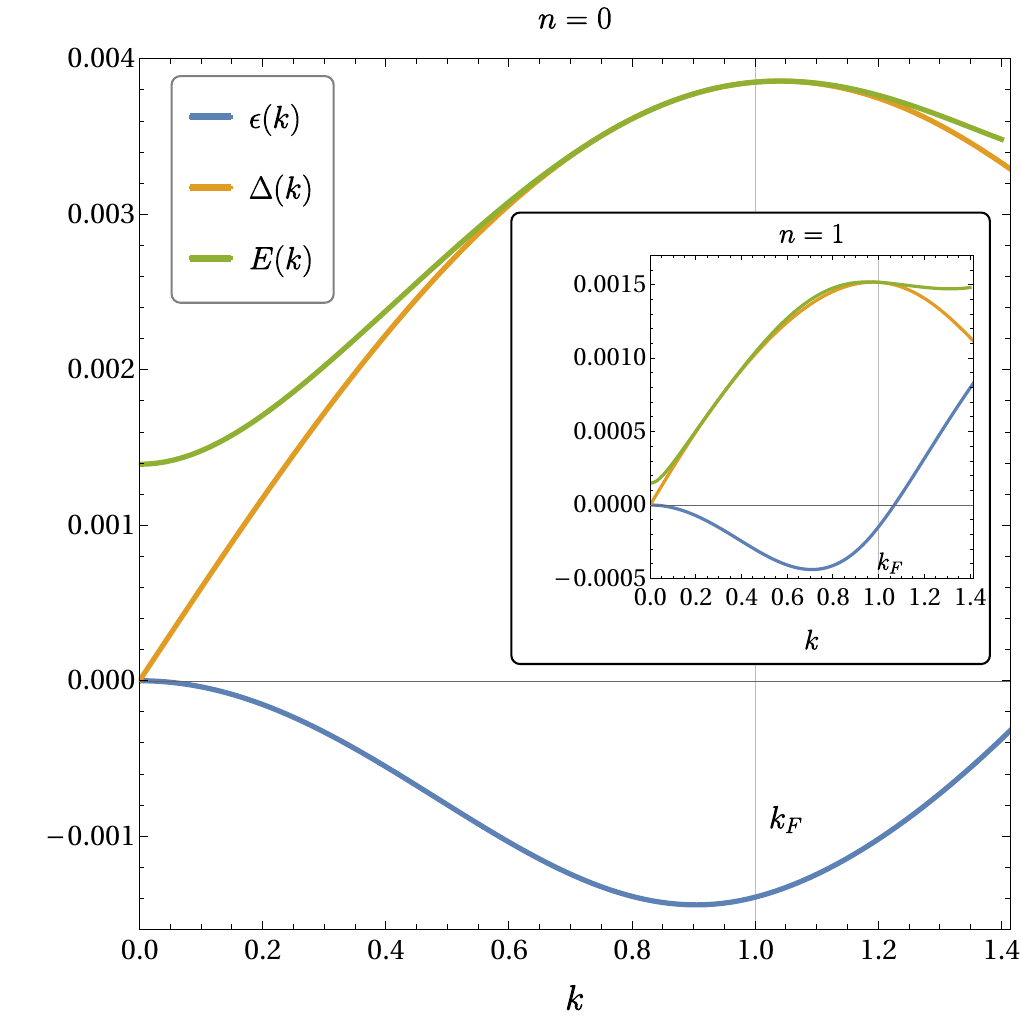}}\\
      \subfloat[\label{cut-off-sb}Pairing solution with $\Lambda_q=\sqrt{2}$]{ \includegraphics[width=0.75\linewidth]{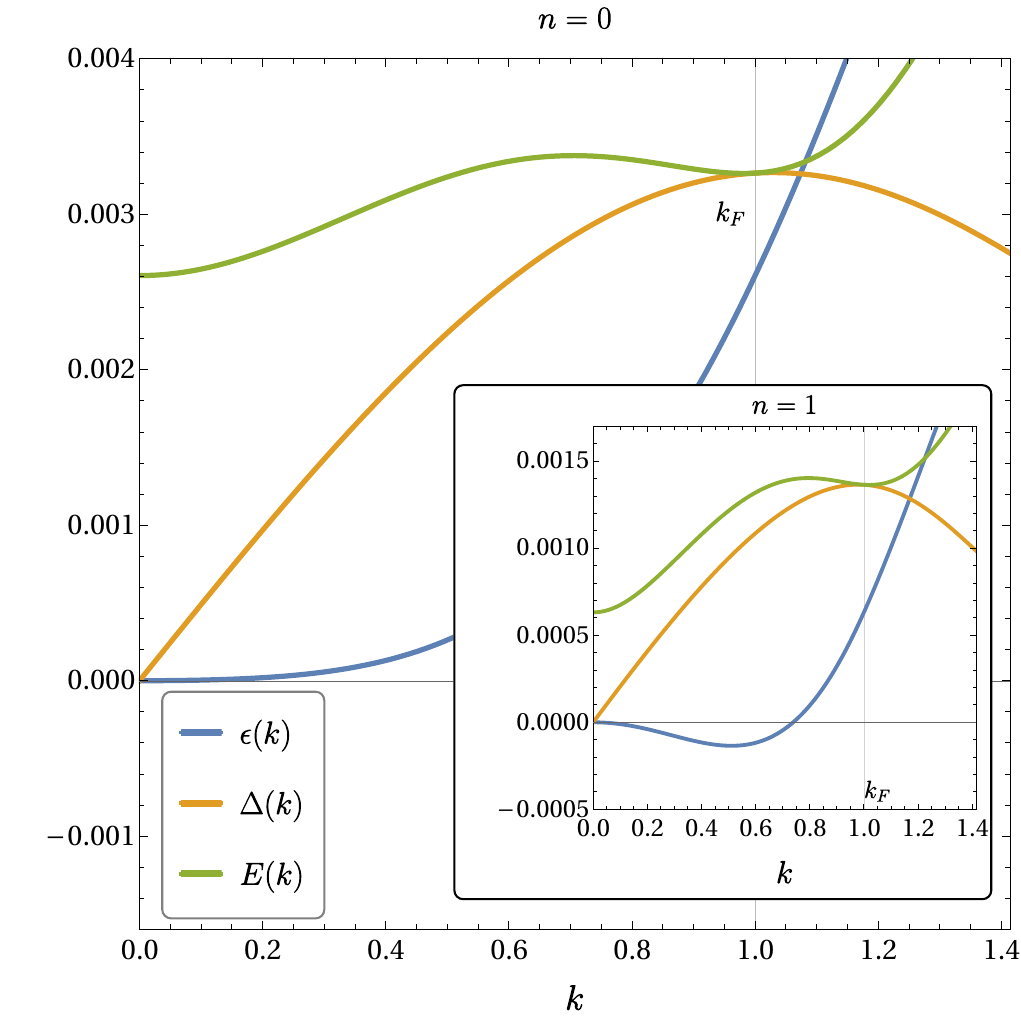}}\\
    \subfloat[\label{cut-off-sc}FL state with $\Lambda_q=\infty$]{ \includegraphics[width=0.75\linewidth]{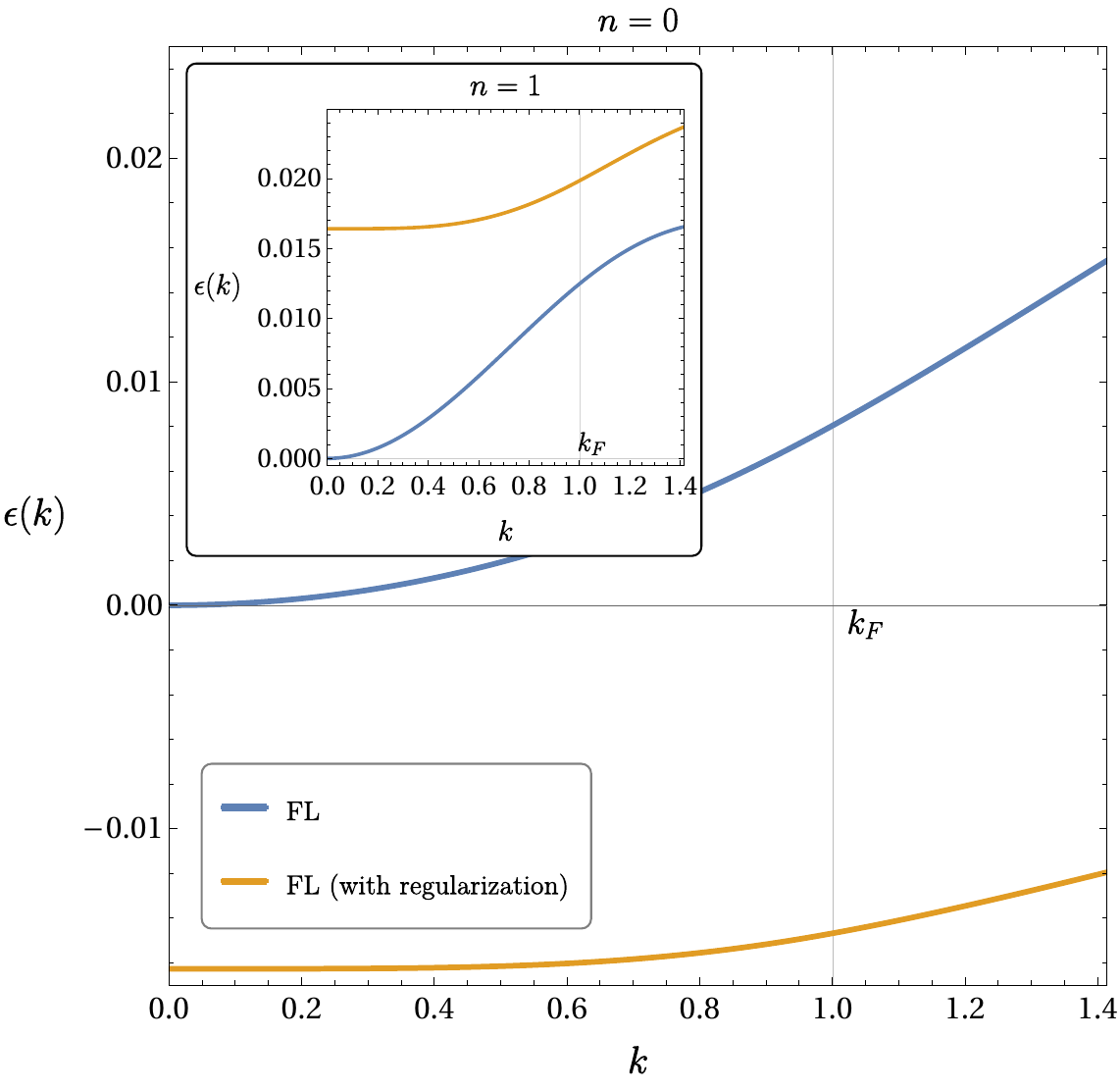}}\\ [-1ex]
   \caption{{Graphics of the relevant variables in the LLL and (insertion) the second LL for two regimes: pairing with cut-offs (a) $\Lambda_q=1/\sqrt{2}$, (b) $\Lambda_q=\sqrt{2}$ and (c) Fermi liquid with $\Lambda_q=\infty$. In all cases $\Lambda_k=\sqrt{2}/l_{B}$ is fixed and the energy is given in unit $V_0=4\pi e^2/l_B$.}}
   \vspace{-0.4cm}
    \label{cut-off-s}
\end{figure}

In the case of FL there is no such length, screening is presumably weak, and we will consider $\Lambda_q=\infty$. Interestingly our conclusions will not change if we introduce a characteristic length ($\Lambda_q=1/(\sqrt{2}l_{B})$) in the case of the second ($n=1$) Landau level. Moreover, in the case of graphene LLs and FL-like states this characteristic length is necessary to conform to the graphene phenomenology in higher LLs \cite{neskovic_unpublished}.

\begingroup   
 \setlength{\tabcolsep}{10pt} 
\renewcommand{\arraystretch}{1.5} 
\begin{table}[h!]\small
\centering
\begin{tabularx}{\linewidth} { 
  || c
  | >{\centering\arraybackslash}X 
  | >{\centering\arraybackslash}X || }
 \hline
  \multicolumn{3}{||c||}{(a) Pairing solution} \\
 \hline\hline
 LL & $\Lambda_q=\sqrt{2}$ & $\Lambda_q=1/\sqrt{2}$ \\ [0.5ex] 
 \hline\hline
 $n=0$ & -0.00017 & -0.00011\\ 
 $n=1$ & -0.000061 & -0.000046 \\
 $n=2$ & -0.000038 & -0.000033 \\ [0ex] 
 \hline
 \end{tabularx}\\\vspace{5pt}
      \begin{tabularx}{\linewidth} { 
  || c 
  | >{\centering\arraybackslash}X 
  | >{\centering\arraybackslash}X || }  
 \hline
  \multicolumn{3}{||c||}{(b) FL state with $\Lambda_q=\infty$ } \\
 \hline\hline
 LL & No regularization & With regularization\\ [0.5ex] 
 \hline\hline
 $n=0$ &0.00036 & -0.0031\\ 
 $n=1$ &0.00058 & 0.0031\\
 $n=2$ &0.00063 & 0.0016\\ [0ex] 
 \hline
 \end{tabularx}\normalsize\\
  \caption{{\label{table} Total energies of (a) pairing solution  and (b) FLL state in different LL for relevant $\Lambda_q$ cut-offs, given in unit $V_0=4\pi e^2/l_B$. Again, in all calculations we set $\Lambda_k=\sqrt{2}/l_{B}$. }}	
    \end{table}
\endgroup
On the other hand, the integrals over momenta directly connected with the momenta of CFs can be bounded in the mean-field  approach by $\sqrt{2}/l_{B}$, i.e., by the value connected with the number of states in a LL. We will denote this cut-off by $\Lambda_k$. Namely, one may expect in a mean-field approach and the effective dipole description, that the enlarged space with each state labeled by two distinct states in the Landau basis, shrinks to the diagonal description in the momentum basis, because we expect that the electron and correlation hole of a dipole will have the same momentum. Therefore $\Lambda_k=\sqrt{2}/l_{B}$. Figure \ref{cut-off-s} summarizes the results with the cut-off-s, whose values we motivated above. Note that in the case in Fig. \ref{cut-off-sa} (due to the deep minima in the single-particle dispersion) the mean field approach is only meaningful for $\Delta\neq0$. (That does not mean that FLL state is ill-defined; its definition and description requires the invariance under the boost ($K$-invariance)). The overall picture that emerges is in accordance with the analysis with the universal cut-off in \ref{An approach with a universal "cut-off"} and the well-known phenomenology; while in the LLL the FL solution dominates, in the second LL the Pfaffian state is well established (see Table \ref{table}).

\section{Conclusions}
In short, inside a LL, an unstable system of special dipoles, of bosonic system at filling factor one or electron system at filling factor one half, at special point(s) in the (two-body) interaction space, can reorganize itself into stable Pfaffian (paired) phases. We described this phenomenon by working with effective Hamiltonians for underlying quasiparticles - CFs that pair. At a special point effective Hamiltonian represents dipole-dipole interaction with an enhanced symmetry, which is broken in the stable, paired phases. The paired states (phases) are strongest in the vicinity of these points, and these critical points act as some kind of sources of the pairing phenomenon. The origin and reason why these critical point(s) exist, is the nature of the quantum space in which bosons or electrons are, together with the special filling requirements that lead to fermionic quasiparticles that often resist ordering into a FL-like state, and therefore pair.

\begin{acknowledgments}
The authors would like to thank I. Herbut for discussions on this and related problems. Numerical computations were performed on the
PARADOX-IV supercomputing facility at the Scientific Computing Laboratory, National Center of Excellence for the Study of
Complex Systems, Institute of Physics Belgrade. We acknowledge funding provided by the Institute of Physics Belgrade, through the grant by the Ministry of Science, Technological Development, and Innovations of the Republic of Serbia.
\end{acknowledgments}

\appendix
\titleformat{\section}[block]{\centering\bfseries}{APPENDIX \thesection:}{0.7em}{\MakeUppercase}
\section{BCS mean-field approach} \label{appendixA}

In this Appendix we apply the BCS mean-field approach to the Hamiltonian
\begin{align}
{\cal H} =\frac{1}{2} \int \frac{d{\bs{q}}}{(2\pi)^2} \;{\tilde V} (|{\bs{q}}|)   
: \rho^{L}({\bs{q}})
 \rho^{L}(-{\bs{q}}):,
\end{align}
with $ {\tilde V} (|{\bs{q}}|) $ denoting a two-body interaction projected to a LL.
By reducing interaction to Cooper's channel and applying the mean field approximation we get to the effective Hamiltonian for the quasiparticles,
\begin{align}\label{RidBCS}
	&K_{\rm eff}=\int\frac{d^2k}{(2\pi)^{2}}\left( \xi_{\bs{k}}c^\dagger_{\bs{k}}c_{\bs{k}}+\frac{1}{2}(\Delta_{\bs{k}}c^\dagger_{\bs{k}} c^\dagger_{-\bs{k}}+\Delta_{\bs{k}}^*c_{-\bs{k}} c_{\bs{k}})\right)  ,\nn\\
 \end{align}
where $\xi_{\bs{k}}=\epsilon_{\bs{k}}-\mu$,  $\epsilon_{\bs{k}}$ is the single-particle kinetic energy and the order parameter is
\begin{equation}\label{orderparRead1}
\Delta_{\bs{k}}=\int\frac{d^2k'}{(2\pi)^{2}}\tilde{V}(\bs{k}-\bs{k}')e^{i\bs{k}\times \bs{k}'} \langle c_{-\bs{k}'} c_{\bs{k}'}\rangle.
\end{equation}
For the complex p-wave pairings, $l=\pm 1$, we take that $\Delta_{\bs{k}}$ is an eigenfunction of angular momentum: $\Delta_{\bs{k}}=|\Delta_{\bs{k}}|e^{il\varphi}$, where $\varphi$ is the polar angle of vector $\bs{k}$. 

Next, we diagonalize the Hamiltonian by introducing canonical transformations 
\begin{equation}
\begin{aligned}
	\alpha_{\bs{k}} &=u_{\bs{k}}c_{\bs{k}}-v_{\bs{k}}c^\dagger_{-\bs{k}},  \\
	 \alpha^\dagger_{-\bs{k}}&=u_{\bs{k}}c^\dagger_{-\bs{k}}+ v^*_{\bs{k}}c_{\bs{k}}, 
\end{aligned}
\end{equation}
so that $\{\alpha_{\bs{k}}, \alpha^\dagger_{\bs{k}}\}=\delta_{\bs{k},\bs{k}' }$ and $|u_{\bs{k}}|^2+|v_{\bs{k}}|^2=1$.
By insisting that $ [\alpha_{\bs{k}},K_{\rm eff}]=E_{\bs{k}}\alpha_{\bs{k}}$ for all $\bs{k}$, which implies that
\begin{equation}
	K_{\rm eff}=\int\frac{d^2k}{(2\pi)^{2}}E_{\bs{k}}\alpha^\dagger_{\bs{k}}\alpha_{\bs{k}}+{\rm const.} ,
\end{equation}
with $E_{\bs{k}}\geq 0$, one obtains (the simplest form of) the
Bogoliubov-de Gennes (BdG) equations,
\begin{equation}\label{BdG}
\begin{aligned}
E_{\bs{k}} u_{\bs{k}}&=\xi_{\bs{k}}u_{\bs{k}}- \Delta^*_{\bs{k}}v_{\bs{k}}\\
E_{\bs{k}}v_{\bs{k}}&=-\xi_{\bs{k}}v_{\bs{k}}-	\Delta_{\bs{k}}u_{\bs{k}}.
\end{aligned}
\end{equation}
This system has non-trivial solutions only for
\begin{equation}
	E_{\bs{k}}=\sqrt{\xi_{\bs{k}}^2+|\Delta_{\bs{k}}|^2}.
\end{equation}

When we rewrite \eqref{orderparRead1} in terms of the Bogoliubov operators and use the relation $u_{\bs{k}}v_{\bs{k}}=-\frac{\Delta_{\bs{k}}}{2E_{\bs{k}}}$, (which follows from BdG) we get
\begin{equation}
		\Delta_{\bs{k}}=-\int\frac{d^2k'}{(2\pi)^{2}}\tilde{V}(\bs{k}-\bs{k}')e^{i\bs{k}\times \bs{k}'} \frac{\Delta_{\bs{k}'}}{2E_{\bs{k}'}}.
\end{equation}
 The two-body interaction is the delta (contact) interaction projected to the LLL,
$ {\tilde V} (|{\bs{q}}|) = V_0  \exp\left(-  \frac{|\bs{q}|^2}{2}\right)$. Also by using that $\Delta_{\bs{k}'}=|\Delta_{\bs{k}'}|e^{il\varphi'}$, where $\varphi'$ is the polar angle of vector $\bs{k}'$, we get an equation that has to be solved self-consistently,
\begin{equation}\label{orderparRead2}
	|\Delta_{\bs{k}}|=-V_0\int\frac{d^2k'}{(2\pi)^{2}}e^{-\frac{|\bs{k}-\bs{k}'|^2}{2}}e^{i\bs{k}\times \bs{k}'} \frac{|\Delta_{\bs{k}'}|e^{il\varphi}}{2E_{\bs{k}'}}.
\end{equation}
First, we can calculate the angular part of integral in \eqref{orderparRead2},
\begin{equation*}
	J_{\varphi}=\int_0^{2\pi}\frac{d\varphi}{2\pi}e^{kq(\cos\varphi-i\sin\varphi)+il\varphi},
\end{equation*}
to get 
\begin{equation}
	J_{\varphi}=\left\{
	\begin{array}{cc}
		0, & l=1 \\
	kk'\geq 0,& l=-1 .
	\end{array} \right. 
\end{equation}
So we can immediately see that in either case $l=\pm 1$ there are no nontrivial and positive solutions of the self-consisting equation \eqref{orderparRead2}.
\section{The critical point and its description}\label{appendixB}

We work with the Hamiltonian
\small
\begin{align}
{\cal H}=\frac{1}{2}\int \frac{d{\bs{q}}}{(2\pi)^2}{\tilde V} (|{\bs{q}}|)
 (\rho^{L}({-\bs{q}}) - \rho^{R}({-\bs{q}}))
  (\rho^{L}({\bs{q}}) - \rho^{R}({\bs{q}})),\nn\\
\end{align}
\normalsize
(together with gauge constraint $\rho_{\bs{q}}^{R} = 0$ on physical space) introduced in \cite{PhysRevB.102.205126}, and we look for a possibility of pairing. We begin by writing this Hamiltonian in a normal ordered form,
\small
\begin{align}\label{DS2}
	{\cal H}=&2\iint\frac{d^2qd^2k}{(2\pi)^{4}}\tilde{V}(|\bs{q}|) \sin^2\left( \frac{\bs{k}\times\bs{q}}{2}\right)c^\dagger_{\bs{k} }c_{\bs{k}}+\\
  +&2\iiint\frac{d^2q d^2k_1d^2k_2}{(2\pi)^{6}}\tilde{V}(|\bs{q}|)\nonumber \\
	& ~~~~~~\times \sin\left( \frac{\bs{k}_1\times\bs{q}}{2}\right) \sin\left( \frac{\bs{k}_2\times\bs{q}}{2}\right) c^\dagger_{\bs{k}_1 +\bs{q}}c^\dagger_{\bs{k}_2 -\bs{q}}c_{\bs{k}_2}c_{\bs{k}_1}.\nonumber
\end{align}
\normalsize
The effective Hamiltonian for the quasiparticles is
\begin{align}
K_{\rm eff}=\int\frac{d^2k}{(2\pi)^{2}}\left( \xi_{\bs{k}}c^\dagger_{\bs{k}}c_{\bs{k}}+\frac{1}{2}(\Delta_{\bs{k}}c^\dagger_{\bs{k}} c^\dagger_{-\bs{k}}+\Delta_{\bs{k}}^*c_{-\bs{k}} c_{\bs{k}})\right),
\end{align}
where order parameter is
\begin{align}
\Delta_{\bs{k}}=-4\int\frac{d^2k'}{(2\pi)^{2}}\tilde{V}(|\bs{k}-\bs{k}'|) \sin^2\left( \frac{\bs{k}\times\bs{k}'}{2}\right)  \langle c_{-\bs{k}'} c_{\bs{k}'}\rangle.
\end{align}
Following the same procedure as in Appendix \ref{appendixA} we end up with the equation that needs to be solved self-consistently,
\small
\begin{align}\label{orderparDS}
|\Delta_{\bs{k}}|=2 V_0\int\frac{d^2k'}{(2\pi)^{2}}e^{-\frac{|\bs{k}-\bs{k}'|^2}{2}}\sin^2\left( \frac{\bs{k}\times\bs{k}'}{2}\right) \frac{|\Delta_{\bs{k}'}|e^{il\varphi'}}{\sqrt{\xi_{\bs{k}'}^2+|\Delta_{\bs{k}'}|^2}}.\nn\\
\end{align}
\normalsize
We can easily carry out the angle integration and for $l=\pm 1$ we have following result, 
\small
\begin{align}\label{orderparDS2}
|\Delta_{\bs{k}}|=&\frac{V_0}{2} e^{-\frac{|\bs{k}|^2}{2}}\int_0^{\infty}\frac{dk'}{2\pi} k' e^{-\frac{|\bs{k}'|^2}{2}}\nn\\
&\times\left(2I_1(kk')-kk' \right)\frac{|\Delta_{\bs{k}'}|}{\sqrt{\xi_{\bs{k}'}^2+|\Delta_{\bs{k}'}|^2}},~~~~~~~
\end{align}
\normalsize
where $I_1(x)$ denote modified Bessel function of the first order. We can see that the BCS mean field approach generates two pairing solutions of opposite angular momenta and thus preserves the Hamiltonian's (extra) symmetry under exchange $\rho^{L}_{\bs{q}} \leftrightarrow \rho^{R}_{\bs{q}}$.

Next, because $I_1(x)\sim \frac{x}{2}$ when $x\rightarrow0$ we can see that order parameter behaves for small $k$ as  $ \Delta_{\bs k} \sim |{\bs k}|^2 (k_x - i k_y )$. According to Ref. \cite{PhysRevB.61.10267} this deviates from behavior that we would expect from a canonical Pfaffian. The  consequences of this anomalous behavior we can see primarily if we examine pairing function, first in momentum space, 
\begin{equation}
g_k=\frac{v_k}{u_k}=-\frac{E_k-\xi_k}{\Delta_k^*}\overset{k\rightarrow0 }{\sim} -\frac{2\mu}{\Delta_k^*}\overset{k\rightarrow0 }{\sim}-\frac{e^{i\phi}}{|{\bs k}|^3}
\end{equation}
and then in coordinate space, 
\small
\begin{align}
g(r)&=\int \frac{d^2k}{(2 \pi)^2}g_ke^{i\bs{k}\cdot\bs{r}}\overset{k\rightarrow0 }{\sim}-\int \frac{d^2k}{(2 \pi)^2}\frac{e^{i\phi}}{k^3}e^{i\bs{k}\cdot\bs{r}}\nn\\
&=-i\lim_{k_1\rightarrow0}\int^{\infty}_{k_1}\frac{dk}{2 \pi}\frac{J_1(kr)}{k^2}
\end{align}
\begin{align}
g(r)&=-\frac{i}{32}\lim_{k_1\rightarrow0}\bigg(k_1^2 r^3 \,_2F_3\left(1,1;2,2,3;-\frac{1}{4} k_1^2 r^2\right)+\nn\\
&\hspace{15mm}+8 r (-2 \log (k_1 r)-2 \gamma +1+\log (4))\bigg)\nn\\
&\hspace{40mm}\text{ if }~k_1>0~\land~ r>0,
\end{align}
\normalsize
where $\,_pF_q  (a;b;z) $ is the generalized hypergeometric function. We can immediately notice that $g_k$ diverges for small $k$, and, as expected, the inverse Fourier transform of $g_k$ also diverge overall when $k_1\rightarrow0$.
But if we focus only on the behavior at long distance,
\begin{align}
g(r)&\xrightarrow{r\rightarrow\infty}-\frac{i}{16\sqrt{2 \pi }}\lim_{k_1\rightarrow0}\frac{\cos\left(k_1 r-\frac{\pi}{4}\right)}{ k_1^{5/2} r^{3/2}}
\overset{\substack{k_1\rightarrow0 \\ r\rightarrow\infty}}{\sim}\frac{1}{k_1^{5/2}r^{3/2}},
\end{align}
the least we can say is that, due to effective pairing $\frac{1}{z|z|}$, this is incompatible with the behavior of allowed wave-functions inside a LL. 

In addition, we may consider the Bogoliubov neutral fermion solutions of the order parameter. The BdG equations \eqref{BdG} for the single vortex when $\bs{k}$ are small and for $E=0$ become 
\begin{equation}\label{BdG2}
\begin{aligned}
\xi_{\bs{k}}u_{\bs{k}}&\approx0,\\
\xi_{\bs{k}}v_{\bs{k}}&\approx0,
\end{aligned}
\end{equation}
because $\Delta_{\bs{k}} \sim |{\bs k}|^2 (k_x - i k_y )$ tends to zero faster then $\epsilon_k\sim |\bs{k}|^2$. In polar coordinates, $\epsilon_k\rightarrow -\frac{1}{2m}\nabla^2$, and  equations \eqref{BdG2} becomes
\begin{equation}\label{BdG3}
\begin{aligned}
-\frac{1}{2m}\left[\frac{1}{r}\frac{\partial}{\partial r} \left(r\frac{\partial}{\partial r}\right)+\frac{1}{r^2}\frac{\partial^2}{\partial \phi^2}\right]u(r)=&\mu v(r)\\
-\frac{1}{2m}\left[\frac{1}{r}\frac{\partial}{\partial r} \left(r\frac{\partial}{\partial r}\right)+\frac{1}{r^2}\frac{\partial^2}{\partial \phi^2}\right]v(r)=&\mu v(r).
\end{aligned}
\end{equation}
The normalizable solution has the form
\begin{equation}
\begin{aligned}
u&=\left(ie^{-i\phi}\right)^{-1/2} f(r)\\
v&= \left(-ie^{i\phi}\right)^{-1/2} f(r)
\end{aligned}
\end{equation}
where $f(r)$ is a real function. Equation \eqref{BdG3} reduces to
\begin{equation}\label{BdG4}
-\frac{1}{2m}\left[\frac{\partial^2}{\partial r^2}+\frac{1}{r}\frac{\partial}{\partial r}-\frac{1}{r^2}\left(\frac{1}{2}\right)^2\right]f(r)=\mu f(r).
\end{equation}
This is a Bessel differential equation, with $k_{\textsc{f}}=\sqrt{2m\mu}$,
\begin{equation}\label{BdG5}
\left[\frac{\partial^2}{\partial r^2}+\frac{1}{r}\frac{\partial}{\partial r}+\left(k_{\textsc f}^2-\frac{\left(\frac{1}{2}\right)^2}{r^2}\right)\right]f(r)=0,
\end{equation}
and the solution, $f(r)=J_{\frac{1}{2}}(k_{\textsc f} r)$, which is not a bound state at zero energy and thus a vortex does not support a Majorana bound state.
 \vfil

\section{Pfaffian states in the mean field approach in the bosonic case}\label{appendixC}

In this Appendix we will demonstrate that Hamiltonian, 
\begin{align}
H^{\rm mod}=&\frac{1}{2} \int \frac{d{^2q}}{(2\pi)^2} {\tilde V} ({{|\bs{q}|}})
: \rho^{L}({\bs{q}})
 \rho^{L}(-{\bs{q}}): + \\
 +&\int \frac{d^2k}{(2\pi)^2} C e^{-\frac{|\bs{q}|^2}{2}}  \rho^{R}({-\bs{q}})
  (\rho^{R}({\bs{q}}) - \rho^{L}({\bs{q}})),\nn
\end{align}
for $C$ large enough, generates Pfaffian states in the mean-field approach. Again, we begin by writing this Hamiltonian in terms of (normally ordered) quasiparticles creation and annihilation operators,
\begin{align}\label{Hmod2}
	H^{\rm mod}=&-iC\int\frac{d^2qd^2k}{(2\pi)^{4}}\tilde{V}(|\bs{q}|) e^{\frac{i}{2}\bs{k}\times\bs{q}} \sin\left( \frac{\bs{k}\times\bs{q}}{2}\right)c^\dagger_{\bs{k} }c_{\bs{k}}+\nonumber\\
  &+\frac{1}{2}\int\frac{d^2q d^2k_1d^2k_2}{(2\pi)^{6}}\tilde{V}(|\bs{q}|)\bigg(e^{-\frac{i}{2}(\bs{k}_1-\bs{k}_2)\times\bs{q}}- \nonumber \\
	&-2iCe^{\frac{i}{2}\bs{k}_1\times\bs{q}}\sin\left( \frac{\bs{k}_2\times\bs{q}}{2}\right)\bigg) c^\dagger_{\bs{k}_1 +\bs{q}}c^\dagger_{\bs{k}_2 -\bs{q}}c_{\bs{k}_2}c_{\bs{k}_1}.\nonumber \\
\end{align}
The effective mean field Hamiltonian is reduced into a quadratic form,
\begin{align}\label{HmodBCS}
	&K_{\rm eff}=\int\frac{d^2k}{(2\pi)^{2}}\left( \xi_{\bs{k}}c^\dagger_{\bs{k}}c_{\bs{k}}+\frac{1}{2}(\Delta_{\bs{k}}c^\dagger_{\bs{k}} c^\dagger_{-\bs{k}}+\Delta_{\bs{k}}^*c_{-\bs{k}} c_{\bs{k}})\right)  ,\nn\\
 \end{align}
where the order parameter is
\begin{align}
	\Delta_{\bs{k}}=&\int\frac{d^2k'}{(2\pi)^{2}}\tilde{V}(|\bs{k}-\bs{k'}|)\big( e^{i\bs{k}\times\bs{k'}}+\nn\\
 &~~~~~~~~~~~~~~+Ce^{-i\bs{k}\times\bs{k'}}-C\big)  \langle c_{-\bs{k'}} c_{\bs{k'}}\rangle.
 \end{align}
Because $\langle c_{-\bs{k'}} c_{\bs{k'}}\rangle= u_{\bs{k'}}v_{\bs{k'}}=-\frac{\Delta_{\bs{k'}}}{2E_{k'}}$ and  $\Delta_{\bs{k'}}=|\Delta_{\bs{k'}}|e^{il\varphi'}$, we end up with the following equation,
\begin{align}
	\Delta_{\bs{k}}=&-\int\frac{d^2k'}{(2\pi)^{2}}\tilde{V}(|\bs{k}-\bs{k'}|)\big( e^{i\bs{k}\times \bs{k'}}+\nn\\
 &~~~~~~~~~~~~~~+Ce^{-i\bs{k}\times\bs{k'}}-C\big)\frac{\Delta_{\bs{k'}}}{2E_{\bs{k'}}},
 \end{align}
with $E_{\bs{k}}=\sqrt{\xi_{\bs{k}}^2+|\Delta_{\bs{k}}|^2}$. Thus, after the angle integration,
\begin{align}\label{selfcon}
	|\Delta_{k}|=&-\frac{V_0}{2}e^{-\frac{k^2}{2}}\int_{0}^{\infty}\frac{dk'}{2\pi}qe^{-\frac{k'^2}{2}}\times\nn\\
 &\times\big(kk'-CI_1(kk') \big)\frac{|\Delta_{\bs{k'}}|}{\sqrt{\xi_{\bs{k'}}^2+|\Delta_{\bs{k'}}|^2}} ,
\end{align}
and again $I_1(x)$ denote modified Bessel function of the first order. There are two important comments; first, the order parameter has a nontrivial solution for Cooper channel $l=-1$ only for $ C>2$, second, in the long-wavelength approximation, $\Delta_{\bs{k}}\sim |\bs{k}|$ which is canonical behavior for the Pfaffian \cite{PhysRevB.61.10267}. 
Next, we need to check that we have sensible dispersion, $\epsilon(\bs{k})$, which increases monotonically as $k$ increases. The dispersion has two contributions. One is from the single particle part of the Hamiltonian \eqref{Hmod2},
\begin{align}
\epsilon_{1}(\bs{k})=\frac{C}{2}\int\frac{d^2q}{(2\pi)^{2}}V_0  e^{-\frac{|\bs{q}|^2}{2}}\left( 1-e^{i\bs{k}\times\bs{q}}\right) ,
\end{align}
where after integration we get
\begin{equation}
\epsilon_{1}(\bs{k})=\frac{CV_0}{4\pi}\left( 1-e^{-\frac{|\bs{k}|^2}{2}}\right).
\end{equation}
The second contribution is the HF term, 
\begin{align}
\epsilon_{\textsc{hf}}(\bs{k})=&\int\frac{d^2k'}{(2\pi)^{2}}\tilde{V}(0)n_{\bs{k'}}-\nn\\
 &-\int\frac{d^2k'}{(2\pi)^{2}}\tilde{V}(\bs{k} -\bs{k'} )\left( 1-Ce^{i\bs{k}\times\bs{k'}}\right) n_{\bs{k'}},
\end{align}
for which we have the following expression, 
\small
\begin{align}\label{eHF}
\epsilon_{\textsc{hf}}(\bs{k})=&V_0 \bigg( \frac{k_{\textsc{f}}}{4\pi}+\frac{C}{2\pi} e^{-\frac{k^2}{2}}  \bigg( 1-e^{-\frac{k_{\textsc{f}}^2}{2}}-\nn\\ 
 &-(1+C)\int_{0}^{k_{\textsc{f}}}\frac{dk'}{2\pi}k'e^{-\frac{k'^2}{2}}I_0(kk')\bigg) \bigg).
\end{align}
\normalsize
Next, we examine Bogoliubov energies, $E_{\bs{k}}=\sqrt{\xi_{\bs{k}}^2+|\Delta_{\bs{k}}|^2}$, as a function of the momentum for different C. By numerically sEqs. \eqref{selfcon} and \eqref{eHF} we get the energies that are plotted in Fig. \ref{FigOrderParBogoliubovDisBoz}.
We can see that the energy of the Bogoliubov excitation for $C\approx2$ has a minimum at $k=k_{\textsc{f}}$, which corresponds to a weak coupling regime. With increasing $C$, we can notice that the minimum is slowly moving from $k=k_{\textsc{f}}$ to $0$. Thus, we find a slow crossover from a weak- to a strong-coupling regime. With this, we conclude that our modified Hamiltonian indeed supports a stable ground state of the Pfaffian kind.


\section{Fermi-liquid description in the electron case}\label{appendixD}

In the following we will describe a regularization procedure on the Hamiltonian in Eq. \eqref{Hameff} in a mean-field approach, that will enable a physical interpretation of the resulting Hamiltonian as the one that describes a Fermi-liquid-like state inside a LL, i.e., possesses the boost (or $K$) invariance.

We begin with the Hamiltonian \eqref{Hameff},
\small
\begin{align}\label{effHam}
   H_{\rm{eff}}=\frac{1}{8}\!\int\!\!\frac{d^2q}{(2 \pi)^2}{\tilde V}(|{\bs{q}}|) &
  \left(\rho^{L}({-\bs{q}})-\rho^{R}({-\bs{q}})\right)\left(\rho^{L}({\bs{q}})-\rho^{R}({\bs{q}})\right),
\end{align}
\normalsize
in which the Coulomb interaction, $ {\tilde V}(|{\bs{q}}|)=\frac{|1}{{|{\bs{q}}|}}e^{-\frac{|{\bs{q}}|^2}{2}}$, is assumed. The Hamiltonian incorporates a dipole symmetry, and commutes with constraint
$
\rho^L ({\bs q}) + \rho^R ({\bs q}) = 0
$
on the physical space.  We demand that the boosting of the entire Fermi sea via  $\bs{k} \rightarrow \bs{k} + \bs{K}$ leaves the energy unchanged, and thus we need an additional constraint that will ensure this boost invariance. One way to implement this is to demand $m^*\rightarrow \infty$, or equivalently  
\widetext
\begin{equation}
 \left(\frac{\partial^2\epsilon(k)}{\partial k^2}\right)_{k=0}= 0.
 \end{equation}
The Hamiltonian \eqref{effHam} have a single particle term $H_1$ and normal ordered one $:H_{\rm{eff}}:$,

 \begin{align}
   H_{\rm{eff}}&=H_1+:H_{\rm{eff}}:\nn\\
  & =\frac{1}{2}\int\frac{d^2q}{(2 \pi)^2}\frac{d^2k}{(2 \pi)^2}       {\tilde V}(|{\bs{q}}|)
 \sin^2{\frac{\bs{k}\times \bs{q}}{2}}  c^\dagger_{\bs{k}} c_{\bs{k}} +
  \frac{1}{2}\int\frac{d^2q}{(2 \pi)^2}\frac{d^2k_1}{(2 \pi)^2}\frac{d^2k_2}{(2 \pi)^2}{\tilde V}(|{\bs{q}}|)
  \sin{\frac{\bs{k_1}\times \bs{q}}{2}}
     \sin{\frac{\bs{k_2}\times \bs{q}}{2}}c^\dagger_{\bs{k_1}+\bs{q}}c^\dagger_{\bs{k_2}-\bs{q}}c_{\bs{k_2}}c_{\bs{k_1}}
 \end{align}

If we interpret the mass in $H_1$ at $k_{\textsc f}$ as the effective mass $m^*$ , of a FL description we can immediately see that 
\begin{equation}
 \frac{1}{m^*}=\left(\frac{\partial^2\epsilon_1(k)}{\partial k^2}\right)_{k=0}\neq 0,
 \end{equation} 
 so we need to regularize our Hamiltonian to achieve that the mass term vanish.

We regularize description by adding a term that represent an interaction, but at the same time it is equal to zero on the physical space \cite{PhysRevB.107.155132}. In the inverse space that may be term of the following form,

\begin{align}
   H_{\rm{add}}(C_{n})&= C_{n}\int \frac{d^2q}{(2 \pi)^2}  e^{-\frac{|{\bs{q}}|^2}{2}}L_n^2\left(\frac{|{\bs{q}}|^2}{2}\right)\left(\rho^{L}({-\bs{q}})+\rho^{R}({-\bs{q}})\right) \left(\rho^{L}({\bs{q}})+\rho^{R}({\bs{q}})\right)\\
    &=4C_{n}\iint\frac{d^2q}{(2 \pi)^2}\frac{d^2k}{(2 \pi)^2} 
    e^{-\frac{|{\bs{q}}|^2}{2}}L_n^2\left(\frac{|{\bs{q}}|^2}{2}\right)\cos^2{\frac{\bs{k}\times \bs{q}}{2}}c^\dagger_{\bs{k}} c_{\bs{k}}+\nn\\
    &+4C_{n}\iiint\frac{d^2q}{(2 \pi)^2}\frac{d^2k_1}{(2 \pi)^2}\frac{d^2k_2}{(2 \pi)^2}  e^{-\frac{|{\bs{q}}|^2}{2}}L_n^2\left(\frac{|{\bs{q}}|^2}{2}\right)
   \cos{\frac{\bs{k_1}\times \bs{q}}{2}}\cos{\frac{\bs{k_2}\times \bs{q}}{2}}c^\dagger_{\bs{k_1}+\bs{q}} c^\dagger_{\bs{k_2}-\bs{q}}c_{\bs{k_2}}c_{\bs{k_1}},\nn
\end{align}
where the coefficients $C_n$ (which are constants independent of $\bs{q}$) are chosen so that overall (regularized) Hamiltonian, 
\begin{align}\label{regHam}
   {\tilde H}&=H_{\rm{eff}}+H_{\rm{add}}(C_{n})\nn\\
   & =\frac{1}{2}\iint\frac{d^2q}{(2 \pi)^2}\frac{d^2k}{(2 \pi)^2}       e^{-\frac{|{\bs{q}}|^2}{2}}L_n^2\left(\frac{|{\bs{q}}|^2}{2}\right) V(|{\bs{q}}|) \sin^2{\frac{\bs{k}\times \bs{q}}{2}}  c^\dagger_{\bs{k}} c_{\bs{k}}+ \nn\\
    &~~~~~~~~~~+\frac{1}{2}\iiint\frac{d^2q}{(2 \pi)^2}\frac{d^2k_1}{(2 \pi)^2}\frac{d^2k_2}{(2 \pi)^2}e^{-\frac{|{\bs{q}}|^2}{2}}L_n^2\left(\frac{|{\bs{q}}|^2}{2}\right) V(|{\bs{q}}|)
    \sin{\frac{\bs{k_1}\times \bs{q}}{2}}\sin{\frac{\bs{k_2}\times \bs{q}}{2}}c^\dagger_{\bs{k_1}+\bs{q}} c^\dagger_{\bs{k_2}-\bs{q}}c_{\bs{k_2}}c_{\bs{k_1}}+\nn\\
     &+4C_{n}\iint\frac{d^2q}{(2 \pi)^2}\frac{d^2k}{(2 \pi)^2} 
    e^{-\frac{|{\bs{q}}|^2}{2}}L_n^2\left(\frac{|{\bs{q}}|^2}{2}\right)\cos^2{\frac{\bs{k}\times \bs{q}}{2}}c^\dagger_{\bs{k}} c_{\bs{k}}+\nn\\
    &~~~~~~~~~~+4C_{n}\iiint\frac{d^2q}{(2 \pi)^2}\frac{d^2k_1}{(2 \pi)^2}\frac{d^2k_2}{(2 \pi)^2}  e^{-\frac{|{\bs{q}}|^2}{2}}L_n^2\left(\frac{|{\bs{q}}|^2}{2}\right)\cos{\frac{\bs{k_1}\times \bs{q}}{2}}\cos{\frac{\bs{k_2}\times \bs{q}}{2}}c^\dagger_{\bs{k_1}+\bs{q}} c^\dagger_{\bs{k_2}-\bs{q}}c_{\bs{k_2}}c_{\bs{k_1}},
\end{align}
has no mass term $\frac{k^2}{2m^*}$, or 
\begin{equation}
 \left(\frac{\partial^2\tilde{\epsilon}_1(k)}{\partial k^2}\right)_{k=0}= 0.
 \end{equation}

\begin{figure*}[t!]
\centering
     \subfloat[No regularization\label{FigPomeranchuka}]{ \includegraphics[width=0.3\linewidth]{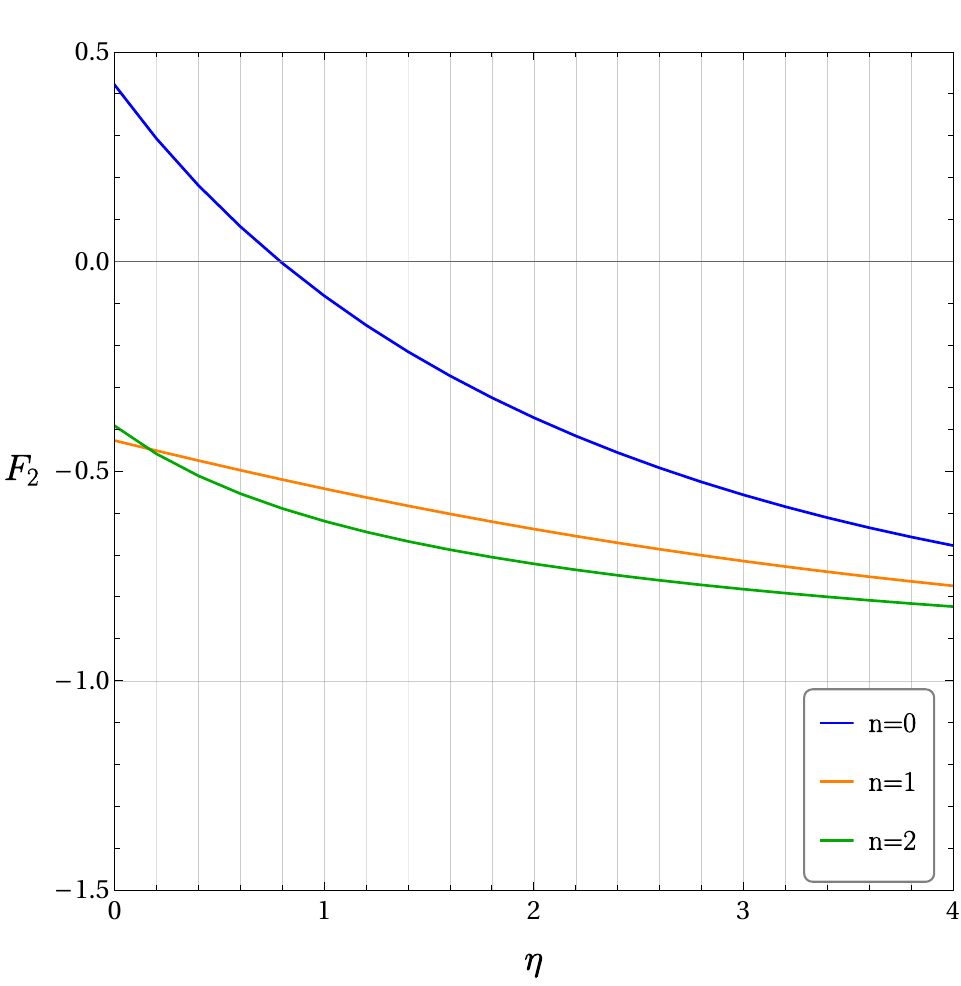}}
      \subfloat[Boost invariant description\label{FigPomeranchukb}]{ \includegraphics[width=0.3\linewidth]{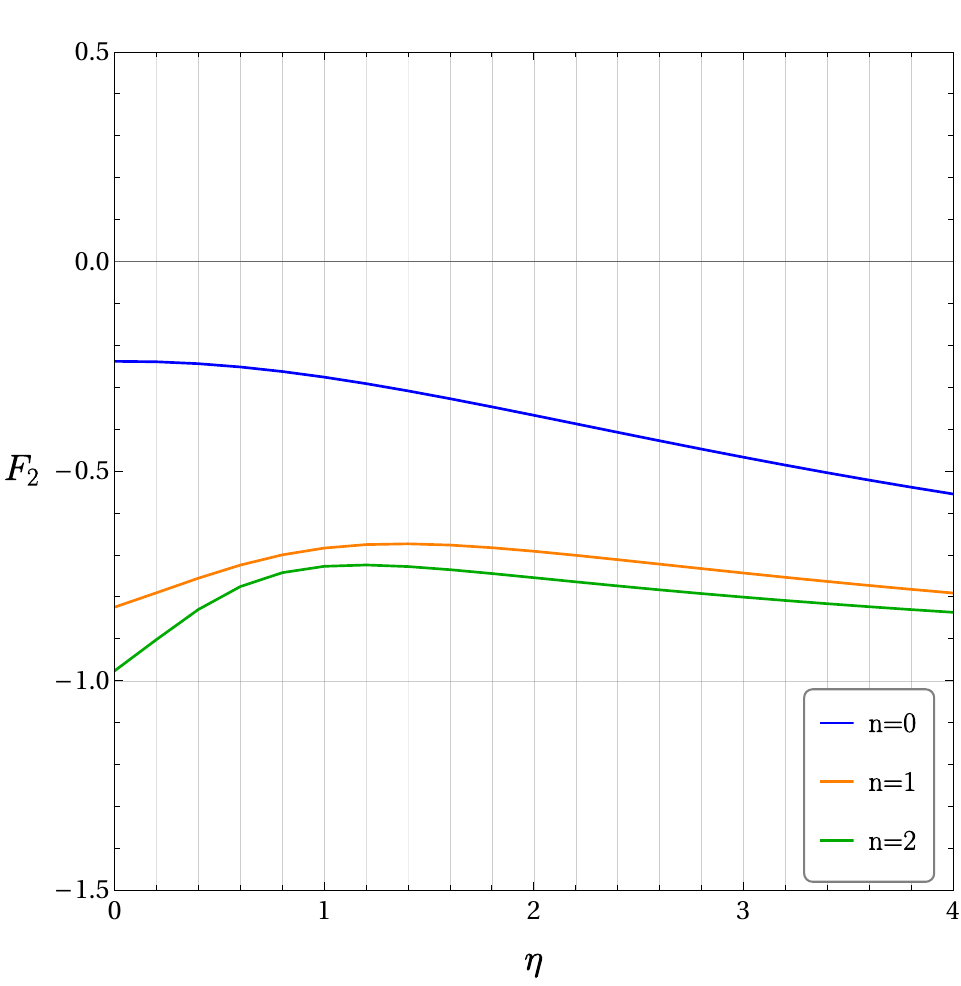}}
       \subfloat[Boost invariant description with introduced cut-off Q=2\label{FigPomeranchukc}]{ \includegraphics[width=0.3\linewidth]{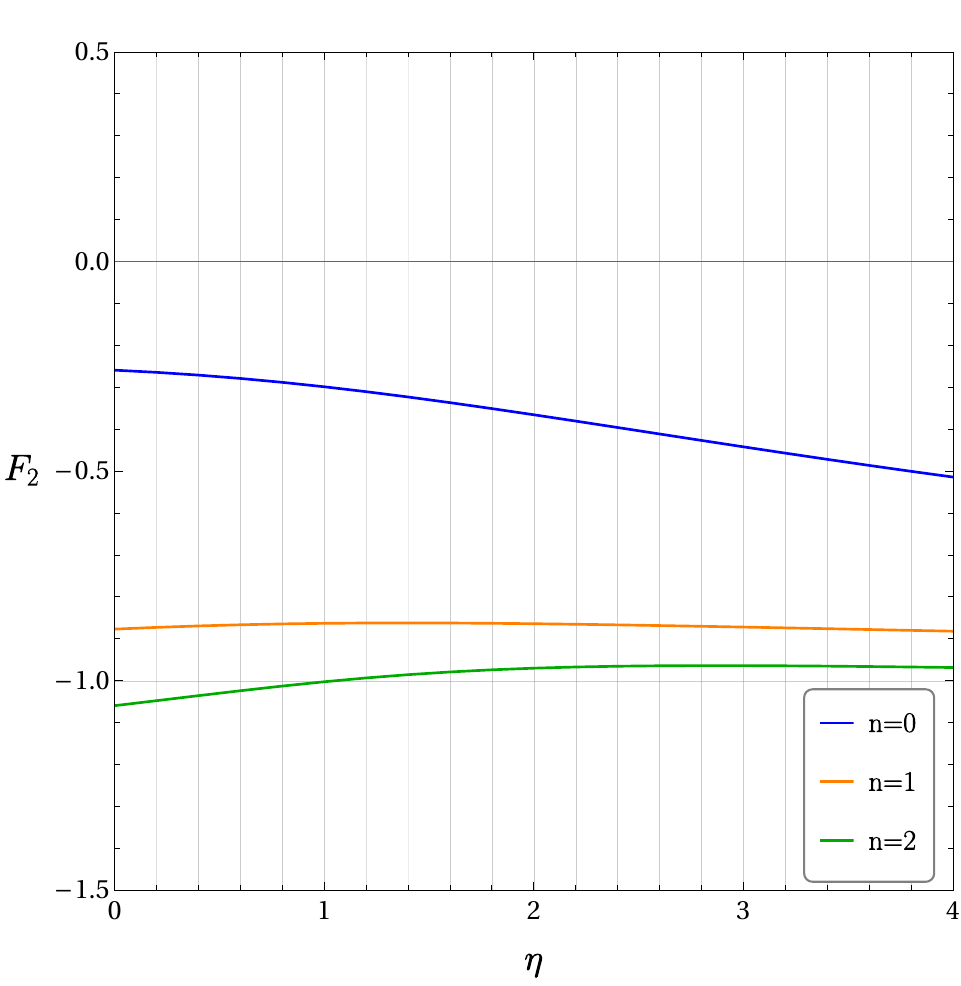}}
   \caption{{Fermi-liquid parameter   $F_2$ for three Landau levels ($n=0,1,2$) plotted as functions of $\eta$}}	
    \label{FigPomeranchuk}
    \vspace{-0.5cm}
\end{figure*}
 
 We get 
\begin{equation}
C_n=\frac{1}{8}\dfrac{\int dq{\tilde V} (|{\bs{q}}|)q^3}{\int dq  e^{-\frac{|{\bs{q}}|^2}{2}}L_n^2\left(\frac{|{\bs{q}}|^2}{2}\right)q^3}=\begin{cases}
\frac{1}{16}\sqrt{\frac{\pi}{2}}, & n=0\\
\frac{7}{192}\sqrt{\frac{\pi}{2}}, & n=1.
\end{cases}
\end{equation}
\newpage
\twocolumngrid
 We can examine expansion of single particle dispersion,
\begin{align}
\epsilon_1(k)=&\frac{1}{2}\int \frac{d^2q}{(2 \pi)^2}e^{-\frac{|{\bs{q}}|^2}{2}}L_n^2\left(\frac{|{\bs{q}}|^2}{2}\right)\bigg( V(|{\bs{q}}|)\sin^2\left(\frac{\bs{k}\times \bs{q}}{2}\right)\nn\\
&+8C_n\cos^2\left(\frac{\bs{k}\times \bs{q}}{2}\right)\bigg)
\end{align}
in terms of $k$:
\begin{itemize}
\item before regularization
\small
\begin{align*}
\text{n=0:}~~&\epsilon_1(k)~\approx~11\cdot 10^{-2} k^2-2\cdot 10^{-3} k^4+3\cdot 10^{-4} k^6\\
&-4\cdot 10^{-5} k^8+3\cdot 10^{-6} k^{10}-2\cdot 10^{-7} k^{12}+2\cdot 10^{-8} k^{14}+\dots,\\
\text{n=1:}~~&\epsilon_1(k)~\approx~2\cdot 10^{-2} k^2-1\cdot 10^{-2} k^4+3\cdot 10^{-3} k^6\\
&-6\cdot 10^{-4} k^8+8\cdot 10^{-5} k^{10}-9\cdot 10^{-6} k^{12}+8\cdot 10^{-7} k^{14}+\dots;
\end{align*}
\normalsize
\item after regularization
\small
\begin{align*}
\text{n=0:}~~&\tilde{\epsilon}_1(k)~\approx~5\cdot 10^{-2}+8\cdot 10^{-4} k^4-2\cdot 10^{-4} k^6\\
&+3\cdot 10^{-5} k^8-3\cdot 10^{-6} k^{10}+3\cdot 10^{-7} k^{12}-2\cdot 10^{-8} k^{14}+\dots,\\
\text{n=1:}~~&\tilde{\epsilon}_1(k)~\approx~3\cdot 10^{-2}+2\cdot 10^{-3} k^4-8\cdot 10^{-4} k^6\\
&+2\cdot 10^{-4} k^8-4\cdot 10^{-5} k^{10}+5\cdot 10^{-6} k^{12}-5\cdot 10^{-7} k^{14}+\dots,
\end{align*}
\normalsize
\end{itemize}
and we can notice that regularized dispersion (1) has no non-analytical terms there, (2) has no $k^2$ term in $\tilde{\epsilon}_1(k)$, as required, and (3) coefficients for $k^i$, $i\geq 4$ are small and decreasing rapidly with increase of power of $k^i$, so we can claim that our regularized Hamiltonian is approximately boost invariant.

To probe the FL description that is governed by our regularized Hamiltonian \eqref{regHam} we consider parameters $F_l$ of the composite fermion Fermi liquid with a generalized Coulomb interaction,
\begin{equation}
V_\eta(q)=\frac{1}{|\boldsymbol{q}|} \exp (-|\boldsymbol{q}| \eta) ,
\end{equation}
which models the effect of the finite-thickness of samples in experiments. The statement that a system is boost invariant is equivalent to saying in a FL description that
\begin{equation}
F_1=-\frac{m^*f_1}{2\pi}=-1,
\end{equation}
and thus
\begin{equation}
F_l=-\frac{m^*f_l}{2\pi}=-\frac{f_l}{f_1}=-\frac{\int d \theta V_{\text {Fock }}^\eta(\theta) \cos l \theta}{\int d \theta V_{\text {Fock }}^\eta(\theta) \cos \theta},
\end{equation}
where 
\begin{align}
V_{\text {Fock }}^\eta(\theta)=& -e^{-1+ \cos\theta}L_n^2\left(-1+ \cos\theta\right)\bigg( V\left(-\sqrt{-1+ \cos\theta}\right)\nn\\
&\times\sin^2\left(\frac{\sin \theta}{2}\right)+8C_n\cos^2\left(\frac{\sin \theta}{2}\right)\bigg).
\end{align}

The calculated $F_2$ as a function of layer width parameter $\eta$, is given in Fig. \ref{FigPomeranchuk} for the first three LLs . We can see that Landau parameter $F_2$ for the regularized Hamiltonian (Fig. \ref{FigPomeranchukb} ) has much better agreement with numerical experiment of Ref. \cite{PhysRevLett.121.147601} than the usual Hamiltonian with dipole symmetry (Fig. \ref{FigPomeranchuka} ). If we introduce cut-off $Q\approx2$ we get that in $n=2$ for some $\eta_c\approx1.2$, $F_2=-1$ as we can see on Fig. \ref{FigPomeranchukc} and this is in agreement with \cite{PhysRevLett.121.147601}, though we do not get the Pomeranchuk instability ($F_2<-1$) in the case of the $n=1$ Landau level at the Coulomb potential ($\eta =0$). 

In the scope of the mean-field approach we will get an additional mass term due to the interactions. This term is the artifact of the method; pure (normal ordered) interaction terms cannot generate a mass, as can be seen in more advanced, quantum Boltzmann equation approach \cite{PhysRevB.107.155132}. Thus, in that respect, the results in Figs. \ref{FigPomeranchukb} and \ref{FigPomeranchukc} should stay the same. If we nevertheless, inside the usual HF approach, try to eliminate the additional mass term (generated by the HF procedure), by selecting appropriate $C$, we have
\begin{equation}
 \left(\frac{\partial^2(\tilde{\epsilon}_1(k)+\tilde{\epsilon}_{\textsc{hf}}(k))}{\partial k^2}\right)_{k=0}= 0.
 \end{equation}
where Hartree-Fock dispersion,

\begin{align}
	\tilde{\epsilon}&_{\textsc{hf}}(\bs{k})=-\int\frac{d^2k'}{(2\pi)^{2}}e^{-\frac{|\bs{k} -\bs{k'} |^2}{2}}L_n^2\left(\frac{|\bs{k} -\bs{k'} |^2}{2}\right)\nn\\
 &\times\left( V(|\bs{k} -\bs{k'} |)\sin^2\left(\frac{\bs{k}\times \bs{k'}}{2}\right)+8C_n\cos^2\left(\frac{\bs{k}\times\bs{k'}}{2}\right)\right)n^{\circ}_{\bs{k'}},
\end{align}
comes from two-particle terms in \eqref{regHam}. We get
\begin{equation}
C_n=\begin{cases}
\frac{2 \sqrt{2 e \pi } \text{erf}\left(\frac{1}{\sqrt{2}}\right)-\sqrt{2 e \pi }-4}{32 \left(\sqrt{e}-1\right)}, & n=0\\
\frac{14 \sqrt{2 e \pi } \text{erf}\left(\frac{1}{\sqrt{2}}\right)-7 \sqrt{2 e \pi }-36}{32 \left(12 \sqrt{e}-29\right)}, & n=1.
\end{cases}
\end{equation} 
where {\rm erf}(x) is the error function. Again, we can check expansions:
\begin{itemize}
\item before regularization on $\epsilon(k)=\epsilon_1(k)+\epsilon_{\textsc{hf}}(k)$
\small
\begin{align*}
\text{n=0:}~~&\epsilon(k)~\approx~8\cdot 10^{-3} k^2+9\cdot 10^{-4} k^4+3\cdot 10^{-1} k^6+3\cdot 10^4 k^8+\dots;\\
\text{n=1:}~~&\epsilon(k)~\approx~2\cdot 10^{-2} k^2-9\cdot 10^{-3} k^4+3\cdot 10^{-1} k^6+3\cdot 10^4 k^8+\dots;
\end{align*}
\normalsize
\item  after regularization on $\epsilon(k)=\epsilon_1(k)+\epsilon_{\textsc{hf}}(k)$
\small
\begin{align*} 
\text{n=0:}~~&\epsilon(k)~\approx~-2\cdot 10^{-2}+2\cdot 10^{-3} k^4+3\cdot 10^{-1} k^6+3\cdot 10^4 k^8+\dots;\\
\text{n=1:}~~&\epsilon(k)~\approx~3\cdot 10^{-2}+6\cdot 10^{-3} k^4+3\cdot 10^{-1} k^6+3\cdot 10^4 k^8+\dots.
\end{align*}
\normalsize
\end{itemize}

The coefficients for the term $k^i$, where $i\geq 6$, are divergent due to an infrared singularity in the integral in the Hartree-Fock term (the coefficients are computed for a certain lower cut-off $Q_1\approx 10^{-3}$). Therefore, even though there is no quadratic term, it is uncertain whether the effective description exhibits the boost invariance.

\section{Many-body phase (CS) transformation between CFs and electrons}\label{appendixE}

In the following we will consider $ V(|{\bs q}|) = -|{\bs q}|^2$, $ {\tilde V}( |{\bs q}|) =  V(|{\bs q}|) \exp(-  |{\bs q}|^2 /2) $ and prove that this two-body interaction between dipoles in $H^{\rm eff}_{\rm Pf}$ correspond, in the long-wavelength expansion, to a leading, three-body contribution \eqref{3bodyelec} that affects elementary electrons.  

Again, like in case of the bosons, in the long-wavelength approximation we may consider  linearized form of $H_{\rm Pf}^{\rm eff}$,
\begin{equation}
H^{\rm eff}_{\rm Pf} \approx
\frac{1}{4}\int \frac{d{\bs{q}}}{(2\pi)^2}{\tilde V}( |{\bs q}|) 
 \rho^{R}({-\bs{q}})
\int \frac{d{\bs{k}}}{(2\pi)^2} (- i {\bs k} \times {\bs q})  c^\dagger_{{\bs k} -{\bs q}} c_{\bs k} .
\end{equation}
The operator $ {\bs K}=\int \frac{d{\bs{k}}}{(2\pi)^2} \;  {\bs k} \;  c^\dagger_{{\bs k} -{\bs q}} c_{\bs k} $ represents the canonical momentum of CFs, and we need to relate this momentum to the one of elementary electrons. If electrons with charge  $q = - e$ are in the presence of the external field, $ {\bs \nabla} \times {\bs A} = - B, B >0$ then Chern-Simons (CS) transformations \cite{ZHANG, PhysRevB.47.7312} relate the mechanical momentum of $i$-th electron  ${\bs \Pi}_e = {\bs p}_i + \frac{e}{c} {\bs A} $ to the mechanical momentum of CF ${\bs \Pi}_{CF} = {\bs p}_i +  \frac{e}{c} {\bs A} - \frac{e}{c}{\bs a} $ for which,
\begin{equation}
{\bs \nabla}_i \times {\bs a} ({\bs x}_i ) = \phi_0 \rho ({\bs x}_i ),\label{nablaa}
\end{equation}
where $\phi_0$ is flux quantum. Thus, in the fermionic field - functional representation, the effective Hamiltonian becomes,
\begin{equation}
{\cal H}^{\rm eff(e)}_{\rm Pf} = \frac{1}{4}\int \frac{d{\bs{q}}}{(2\pi)^2} \left(-|{\bs q}|^2\right) \;  \label{elecfHam}
 \rho ({-\bs{q}})
  (- i )\; ({\bs K} + \frac{e}{c}{\bs a}) ({\bs q})  \times {\bs q} ,
\end{equation}
where $  \rho ({\bs q})$ is the unprojected density of electrons (the same as of CFs). Again, vortex density may be neglected and we are left with the following term,
\begin{align}
{\cal H}^{\rm eff(e)}_{\rm Pf}&  \approx
-\frac{e}{4c}\int \frac{d{\bs{q}}}{(2\pi)^2} |{\bs q}|^2   \rho ({-\bs{q}}) \int \frac{d{\bs{k}}}{(2\pi)^2}\nn\\
 & \hspace{1cm}\times(-i)  \rho ({-\bs{k}})  ({\bs a} ({\bs k} + {\bs q})  \times {\bs q})  \label{elecfHam2}\\
& = \frac{ie V_0 }{4c}\int \frac{d{\bs{q}}}{(2\pi)^2}\; \int \frac{d{\bs{k}}}{(2\pi)^2}\;\frac{|{\bs q}|^2+|{\bs k}|^2}{2}\nn\\
 & \hspace{1cm}\times\rho ({-\bs{q}})  \rho ({-\bs{k}})  \left({\bs a} ({\bs k} + {\bs q})  \times \frac{{\bs q}+{\bs k}}{2}\right)+ \label{elecfHam3}\\ 
&   +\frac{ie V_0 }{4c}\int \frac{d{\bs{q}}}{(2\pi)^2}\; \int \frac{d{\bs{k}}}{(2\pi)^2}\; \frac{|{\bs q}|^2-|{\bs k}|^2}{2}\nn\\
 & \hspace{1cm}\times\rho ({-\bs{q}})  \rho ({-\bs{k}})  \left({\bs a} ({\bs k} + {\bs q})  \times \frac{{\bs q}-{\bs k}}{2}\right). \label{elecfHam4}
\end{align}
It can be shown that the term in \eqref{elecfHam3} has no effect on electrons (because it is equivalent to $\sim (\nabla^2\delta)\delta(x)$). Next, following from \eqref{nablaa}, 
\begin{equation}
{\bs a} ({\bs k} + {\bs q})  \times \frac{{\bs q}-{\bs k}}{2}=\frac{i\phi_0}{4}\frac{|{\bs q}|^2-|{\bs k}|^2}{|{\bs q}+{\bs k}|^2}\rho ({\bs{q}}+{\bs{k}})
\end{equation}
and inserting this to \eqref{elecfHam4} we get
\begin{align}
{\cal H}^{\rm eff(e)}_{\rm Pf}  &\approx -\frac{e \phi_0 V_0 }{32c}\int \frac{d{\bs{q}}}{(2\pi)^2}\; \int \frac{d{\bs{k}}}{(2\pi)^2}\; \frac{\left(|{\bs q}|^2-|{\bs k}|^2\right)^2}{|{\bs q}+{\bs k}|^2}\nn\\
 & \hspace{1cm}\times\rho ({\bs{q}}+{\bs{k}})  \rho ({-\bs{k}})  \rho ({-\bs{q}})  \rho ({-\bs{k}}).\label{elecfHam5}
\end{align}
Shifting ${\bs k}\rightarrow{\bs k}-{\bs q}$ and using l’Hopital’s rule for multivariable functions we end up with the following expression, 
\begin{align}
{\cal H}^{\rm eff(e)}_{\rm Pf}  \approx -\frac{e \phi_0 V_0 }{16c}&\int \frac{d{\bs{q}}}{(2\pi)^2} \int \frac{d{\bs{k}}}{(2\pi)^2} \left(3|{\bs k}-{\bs q}|^2-|{\bs q}|^2\right)\nn\\
&\times\rho ({\bs{k}})  \rho ({-\bs{q}})  \rho ({\bs{q}}{-\bs{k}}).\label{elecfHam6}
\end{align}
We drop term with $|{\bs q}|^2$ due to no effect on electrons. Introducing the Fourier transform on densities, \eqref{elecfHam6} becomes
\begin{align}
{\cal H}^{\rm eff(e)}_{\rm Pf}  &\approx   \frac{3e\phi_0 V_0 }{16c}\int \frac{d{\bs{q}}}{(2\pi)^2}  \frac{d{\bs{k}}}{(2\pi)^2} \int d{\bs x}_1  d{\bs x}_2  d{\bs x}_3  
  \left(-|{\bs k}-{\bs q}|^2\right)\nn\\
  &~~~~\times e^{- i (\bs{q}-\bs{k}) \cdot \bs{x}_1 } 
  e^{- i \bs{k} \cdot \bs{x}_2 }    e^{ i  \bs{q} \cdot \bs{x}_3 }\rho (\bs{x}_1)  \rho (\bs{x}_2)   \rho(\bs{x}_3)\nn\\
&= \frac{3e\phi_0 V_0 }{16c} \int d{\bs x}_1   d{\bs x}_2  d{\bs x}_3  \nabla^2_1\big(\delta({\bs x}_1-{\bs x}_2)\nn\\
&~~~~\times\delta({\bs x}_1-{\bs x}_3)\big)\rho (\bs{x}_1)  \rho (\bs{x}_2)  \rho(\bs{x}_3).
\end{align}
\twocolumngrid

\twocolumngrid
\bibliography{RefTopPair.bib}

\end{document}